\documentclass[twocolumn]{myaastex631} 
\usepackage{graphicx,url}
\usepackage{amsmath,amssymb}
\usepackage{hyperref}
\hypersetup{colorlinks,
	linkcolor=blue, 
	urlcolor=magenta, 
	anchorcolor=blue,
	citecolor=blue
}
\usepackage{bm}
\usepackage{xcolor}
\usepackage{mathrsfs}
\usepackage{natbib}
\usepackage{soul} 
\newcommand{\norm}[1]{\left\lVert#1\right\rVert}

\newcommand\blfootnotetext[1]{%
	\begingroup
	\renewcommand\thefootnote{}\footnotetext{#1}%
	\addtocounter{footnote}{-1}%
	\endgroup}

\definecolor{greenW}{rgb}{0.0, 0.55, 0.1}
\definecolor{orangeW}{rgb}{1.0, 0.5, 0.05}

\begin{document}
\title{\large\bfseries Likelihood-free Inference with Mixture Density Network}

\author{Guo-Jian Wang}
\author{Cheng Cheng}
\author{Yin-Zhe Ma$^{\dagger}$}
\blfootnotetext{$^{\dagger}$ Corresponding author: ma@ukzn.ac.za}
\affil{School of Chemistry and Physics, University of KwaZulu-Natal, Westville Campus, Private Bag X54001, Durban, 4000, South Africa}
\affil{NAOC-UKZN Computational Astrophysics Centre (NUCAC), University of KwaZulu-Natal, Durban, 4000, South Africa}
\affil{National Institute for Theoretical and Computational Sciences (NITheCS), South Africa}

\author{Jun-Qing Xia}
\affil{Department of Astronomy, Beijing Normal University, Beijing 100875, China}

\received{2022 March 20}
\revised{2022 June 28}
\accepted{2022 June 29}
\published{2022 September 2}

\begin{abstract}
In this work, we propose using the mixture density network (MDN) to estimate cosmological parameters. We test the MDN method by constraining parameters of the $\Lambda$CDM and $w$CDM models using Type Ia supernovae and the power spectra of the cosmic microwave background. We find that the MDN method can achieve the same level of accuracy as the Markov Chain Monte Carlo method, with a slight difference of $\mathcal{O}(10^{-2}\sigma)$. Furthermore, the MDN method can provide accurate parameter estimates with $\mathcal{O}(10^3)$ forward simulation samples, which are useful for complex and resource-consuming cosmological models. This method can process either one data set or multiple data sets to achieve joint constraints on parameters, extendable for any parameter estimation of complicated models in a wider scientific field. Thus, the MDN provides an alternative way for likelihood-free inference of parameters.
\end{abstract}
\keywords{Cosmological parameters (339); Observational cosmology (1146); Computational methods (1965); Astronomy data analysis (1858); Neural networks (1933)}

\section{\bf Introduction}\label{sec:introduction}

In research on cosmology, precise parameter estimation is one of the most important steps to understanding the physical processes in the universe. The most commonly used method for parameter estimation is the Markov Chain Monte Carlo (MCMC) method. However, we often encounter scenarios where although we can generate simulated data through forward simulations, the likelihood is intractable in practice. In addition, simulations may consume a lot of time and computational resources, making parameter inference unpragmatic---for example, for physics related to nonlinear structure formation on small scales \citep{Springel:2005,Klypin:2011} and the epoch of reionization \citep{Mesinger:2016,Kern:2017}. Therefore, we need to develop new methods to circumvent these problems.

Recently, the artificial neural network (ANN) with many hidden layers has achieved significant improvement in many fields including cosmology and astrophysics. For example, it performs excellently in searching for and estimating the parameters of strong gravitational lenses \citep{Hezaveh:2017,Jacobs:2017,Petrillo:2017,Pourrahmani:2018,Schaefer:2018,LiRui:2020,LiRui:2021}, analyzing gravitational waves \citep{George:2018a,George:2018b,George:2018c,Shen:2019,LiXiangru:2020}, classifying the large-scale structure of the universe \citep{Aragon-Calvo:2019}, discriminating between cosmological and reionization models \citep{Schmelzle:2017,Hassan:2018}, recovering the cosmic microwave background (CMB) signal from contaminated observations \citep{Petroff:2020,Casas:2022,Wanggj:2022}, reconstructing functions from observational data \citep{Wanggj:2020b,Escamilla-Rivera:2020,Wanggj:2021}, and even estimating cosmological and astrophysical parameters \citep{Shimabukuro:2017,Fluri:2018,Schmit:2018,Fluri:2019,Ribli:2019,Wanggj:2020a,Ntampaka:2020,Nygaard:2022}.

As an application of the ANN, the mixture density network (MDN; \citealt{Bishop:1994}) is designed to model the conditional probability density $p(\bm\theta|\bm{d})$. The MDN is a combination of the ANN and mixture model, which in principle represents an arbitrary conditional probability density. Much of recent literature has used the MDN to model parameters \citep{Zen:2014,Alsing:2018,Alsing:2019,Kruse:2020,Zhao:2022}. In \citet{Alsing:2018}, the MDN was used to model the joint probability density $p(\bm\theta, \bm{t})$, where $\bm{t}$ is the summary statistics of the data. Then, the posterior distribution was obtained by evaluating the joint probability density at the observational summary $\bm{t}_0$, $p(\bm\theta|\bm{t}_0)\propto p(\bm\theta, \bm{t}{\rm =}\bm{t}_0)$. In comparison, in \citet{Alsing:2019}, the MDN was used to model the conditional probability density $p(\bm{t}|\bm\theta)$, and then the likelihood $p(\bm{t}_0|\bm\theta)$ could be obtained at the observational summary $\bm{t}_0$. The posterior distribution could be obtained by multiplying the likelihood and the prior: $p(\bm\theta|\bm{t}_0)\propto p(\bm{t}_0|\bm\theta)\times p(\bm\theta)$.

In this work, we show that for one-dimensional observational data, the MDN is capable of estimating cosmological parameters with high accuracy. The MDN used here aims to model the conditional probability density $p(\bm\theta|\bm{d})$, and the posterior distribution can be obtained at the observational data $\bm{d}_0$. Two mixture models are considered in our analysis: the Gaussian mixture model and the beta mixture model. We test the MDN method by estimating parameters of the $\Lambda$CDM and $w$CDM cosmologies using Type Ia supernovae (SN-Ia) and the angular power spectra of the CMB. The code and examples are available online.\footnote{\url{https://github.com/Guo-Jian-Wang/mdncoper}}

This paper is organized as follows: In Section \ref{sec:methodology}, we illustrate the method of estimating parameters using the MDN, which includes an introduction to the ANN, mixture model, MDN, and training and parameter inference method. Section \ref{sec:apply_to_panthon} shows the application of the MDN method to the Pantheon SN-Ia. Section \ref{sec:joint_constraint_on_parameters} presents a joint constraint on parameters using the power spectra of the {\it Planck} CMB and the Pantheon SN-Ia. Section \ref{sec:effect_of_hyperparameters} shows the effect of hyperparameters of the MDN on the parameter inference. In Section \ref{sec:beta_mixture_model}, we illustrate the beta mixture model. A discussion of the MDN method is presented in Section \ref{sec:discussions}. We then conclude in Section \ref{sec:conclusions}.

\begin{figure}
	\centering
	\includegraphics[width=0.45\textwidth]{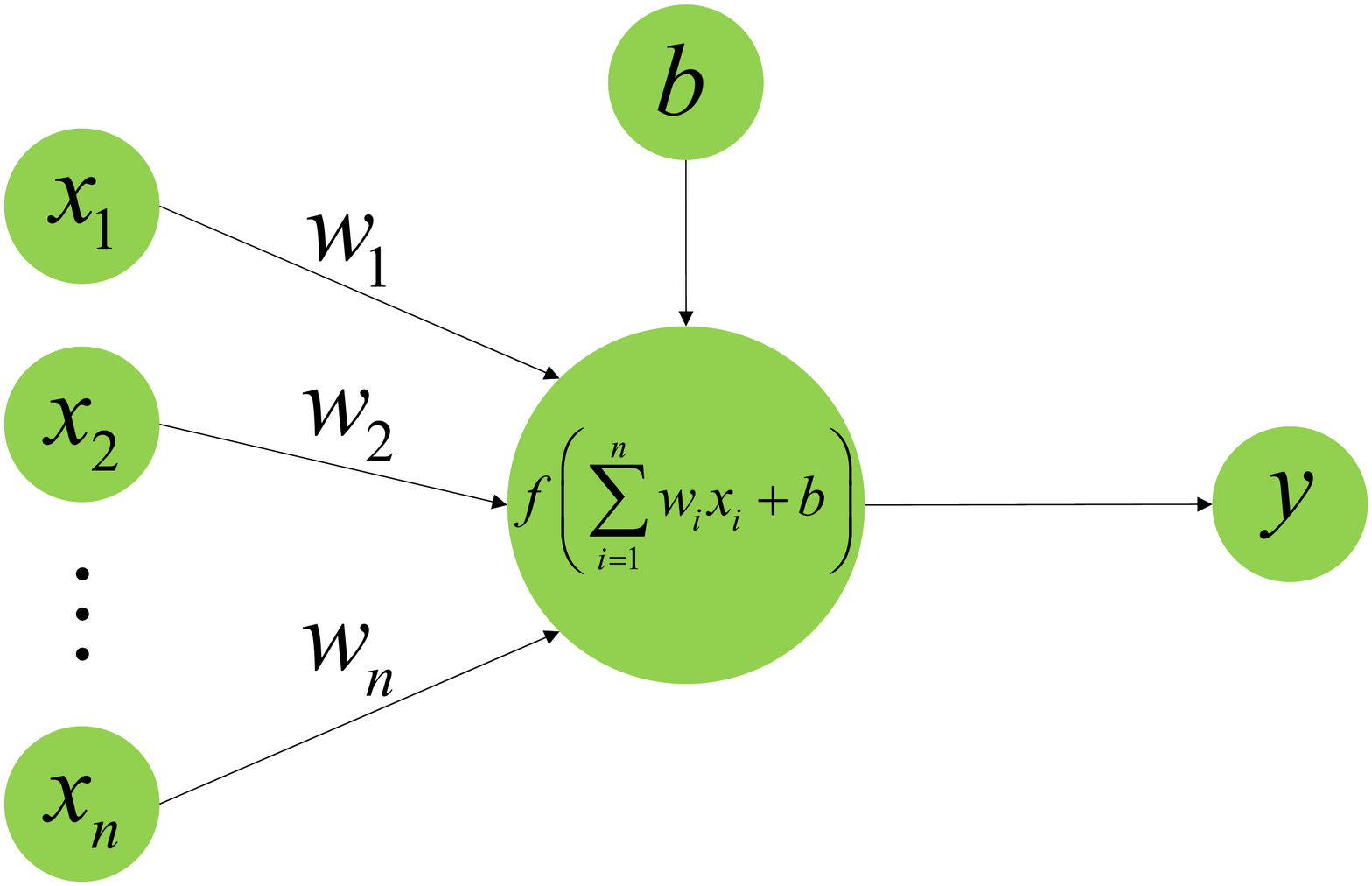}
	\caption{A neuron of the ANN.}\label{fig:neuron}
\end{figure}

\section{\bf Methodology}\label{sec:methodology}

\subsection{ANNs}\label{sec:ann}

An ANN is a computing system inspired by the biological neural network. The unit of an ANN is called a neuron (see Figure~\ref{fig:neuron}), and an ANN consists of many neurons. Each neuron transforms the input from other neurons and gives an output
\begin{equation}\label{equ:neuron_function}
y = f\left(\sum_i w_i x_i + b\right)~,
\end{equation}
where $x$ is the input of a neuron, $f(\cdot)$ is a nonlinear function that is usually called the activation function, and $w$ and $b$ are parameters to be learned by the network. The parameters of an ANN model can be optimized by minimizing the loss function $\mathcal{L}$ in the training process. The loss function used in this work will be described in Section \ref{sec:mdn}. In our analysis, we consider the randomized leaky rectified linear unit (RReLU;~\citealt{RReLU}) as the activation function:
\begin{equation}
f(x) = \left\{\begin{matrix}
x & \text{if } x \geq 0 \\
ax & \text{if } x < 0,
\end{matrix}\right.
\end{equation}
where $a$ is a random number sampled from a uniform distribution $U(l, u)$, where $l, u \in [0, 1)$. Here we adopt the default settings of $l=1/8$ and $u=1/3$ in PyTorch\footnote{\url{https://pytorch.org/docs/stable/index.html}}. A discussion about the effect of activation functions on the result will be shown in Section \ref{sec:effect_of_activationFunction}.

The general structure of an ANN is composed of many layers, and each layer contains many neurons. There are three types of layers in a network: an input layer, one or more hidden layers, and an output layer. Data is fed to the input layer, and then the information is passed through each hidden layer and finally computed from the output layer. Furthermore, the batch normalization technique~\citep{Ioffe:2015} is applied before each nonlinear layer to facilitate optimization and speedup convergence.

In general, there is no theory to determine exactly how many neurons should be used in each hidden layer. Here we adopt the model architecture used in \citet{Wanggj:2020a}, i.e. the number of neurons in each hidden layer is decreased in proportion to the layer index. The number of neurons in the $i$th hidden layer is therefore
\begin{equation}
N_i = \frac{N_{\text{in}}}{F^i}~,
\end{equation}
where $N_{\text{in}}$ is the number of neurons in the input layer and $F$ is a decreasing factor defined by
\begin{equation}
F = \left(\frac{N_{\text{in}}}{N_{\text{out}}} \right)^{\frac{1}{n+1}}~,
\end{equation}
where $N_{\text{out}}$ is the number of neurons in the output layer and $n$ is the number of hidden layers. In our analysis, we consider a network with three hidden layers. We discuss the effect of the number of hidden layers on the result of estimation in Section~\ref{sec:effect_of_hiddenLayer}.

\begin{figure}
	\centering
	\includegraphics[width=0.45\textwidth]{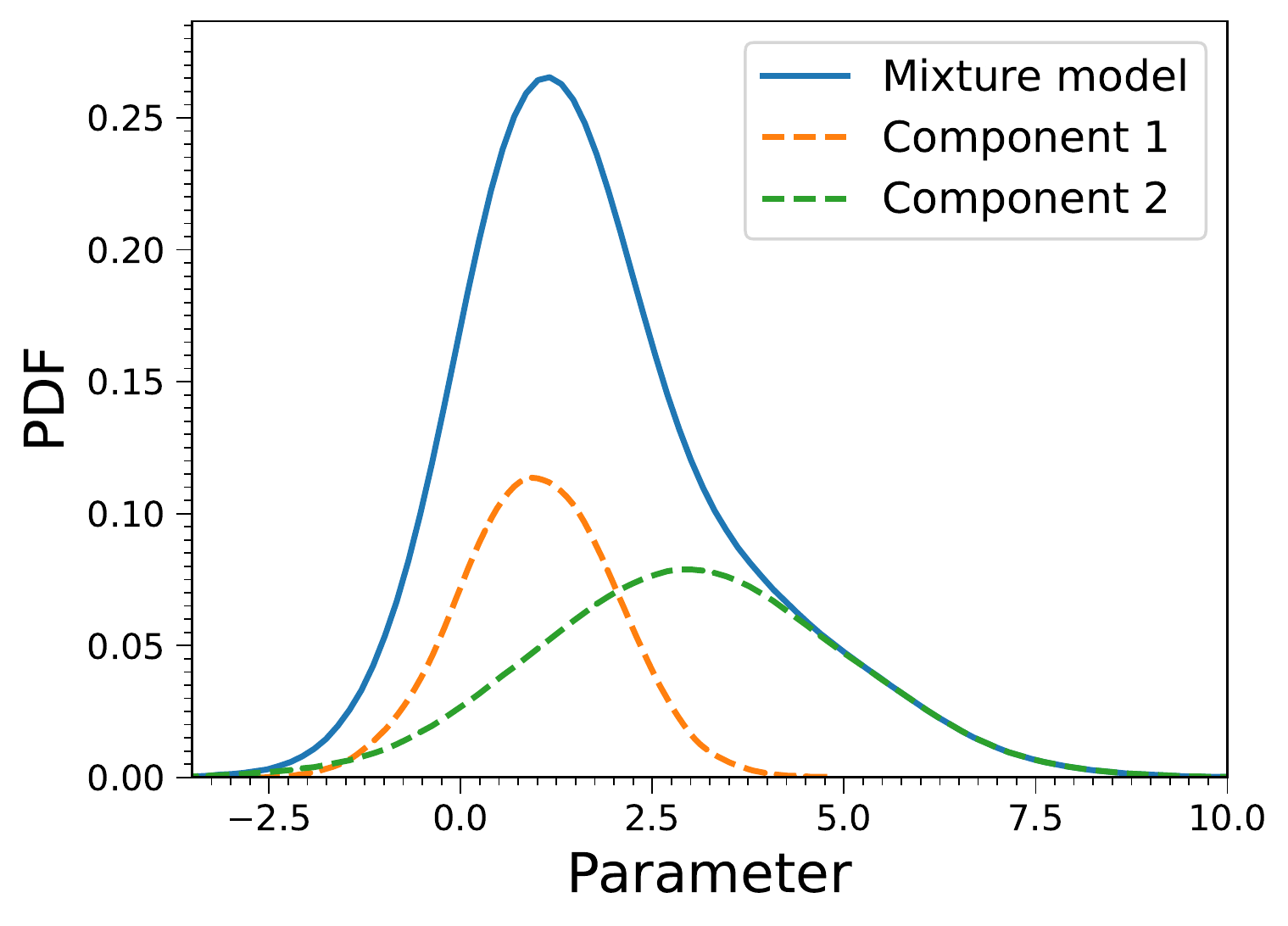}
	\caption{An example of a mixture model.}\label{fig:mixture_model_example}
\end{figure}

\subsection{Mixture Model}\label{sec:gaussian_mixture_model}

A mixture model is a probabilistic model that assumes all data points are generated from a mixture of a finite number of distributions with unknown parameters, where the distribution can be of any type. In Figure~\ref{fig:mixture_model_example} we show an example of a mixture model of a parameter. For measurement $\bm{d}$ and parameters $\bm\theta$, the probability density of $\bm\theta$ with $K$ components has the form
\begin{equation}\label{equ:pdf_of_mixture_model}
p(\bm\theta|\bm{d}) = \sum_{i=1}^K \omega_i p_i(\bm\theta|\bm{d})~,
\end{equation} 
where $\omega_i$ is a mixture weight representing the probability that $\bm\theta$ belongs to the $i$th component. The mixture weights are nonnegative and their sum is normalized, i.e. $\sum_{i=1}^{K}\omega_i = 1$.

The normal (or Gaussian) distribution acts as the foundation for many modeling approaches in statistics. Therefore, in the following analysis, we mainly consider the Gaussian distribution as the unit of the mixture model. In Section~\ref{sec:beta_mixture_model}, we also discuss the applications of the beta mixture model in the parameter estimation. Therefore, for the case of only one parameter, Equation~(\ref{equ:pdf_of_mixture_model}) becomes
\begin{align}\label{equ:pdf_of_gaussian_1}
\nonumber p(\theta|\bm{d}) &= \sum_{i=1}^K \omega_i\mathcal{N}(\theta; \mu_i, \sigma_i) \\
&= \sum_{i=1}^K \omega_i\cdot\frac{1}{\sqrt{2\pi\sigma^2_i}}e^{-\frac{(\theta-\mu_i)^2}{2\sigma^2_i}}~.
\end{align}
For multiple parameters, Equation~(\ref{equ:pdf_of_mixture_model}) becomes
\begin{align}\label{equ:pdf_of_gaussian_multi}
\nonumber p(\bm\theta|\bm{d}) &= \sum_{i=1}^K \omega_i\mathcal{N}(\bm\theta; \bm\mu_i, \bm\Sigma_i) \\
&= \sum_{i=1}^K \omega_i\cdot\frac{\exp{\left( -\tfrac{1}{2} (\bm\theta - \bm\mu_i)^\top \bm\Sigma_i^{-1} (\bm\theta - \bm\mu_i) \right)}}{\sqrt{\left( 2\pi \right)^N |\bm\Sigma_i|}}~,
\end{align}
where $N$ is the number of parameters.

\subsection{MDN}\label{sec:mdn}

\begin{figure}
	\centering
	\includegraphics[width=0.45\textwidth]{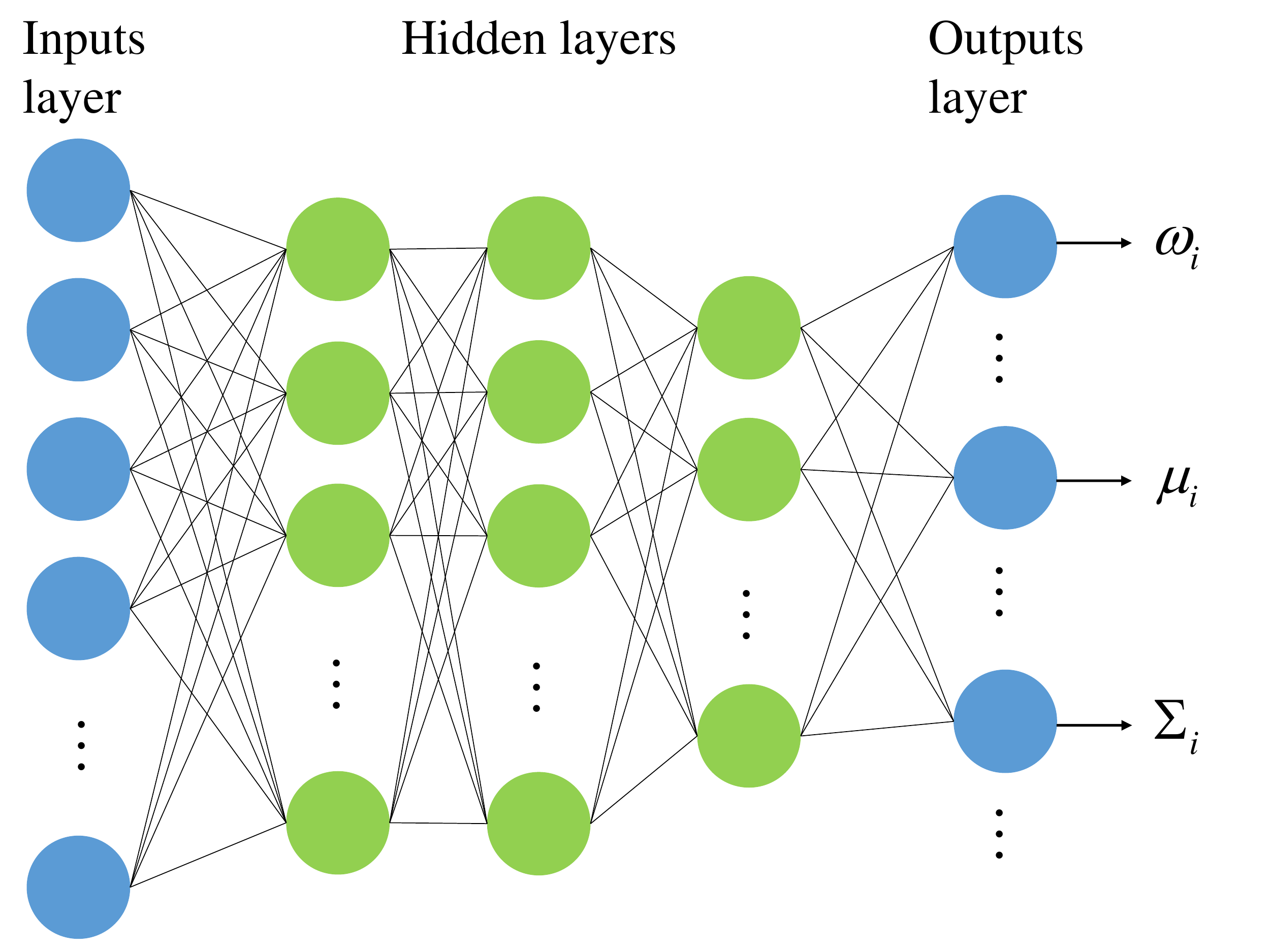}
	\caption{The general structure of an MDN with Gaussian components. The input is the observational data, and the outputs are the parameters of the mixture model. Each node here is a neuron.}\label{fig:mdn}
\end{figure}

An MDN is a combination of an ANN and a mixture model. It learns a mapping between the measurement $\bm{d}$ and parameters of the mixture model. For an MDN with Gaussian components specifically, it learns a mapping between the measurement $\bm{d}$ and the parameters of the Gaussian mixture model ($\omega$, $\bm\mu$, and $\bm\Sigma$ (or $\sigma$ for the case of only one parameter)). Figure \ref{fig:mdn} shows the general structure of an MDN with Gaussian components. The left side of the MDN accepts the measurement as input and then the parameters of the mixture model are computed from the right side.

\begin{figure}
	\centering
	\includegraphics[width=0.45\textwidth]{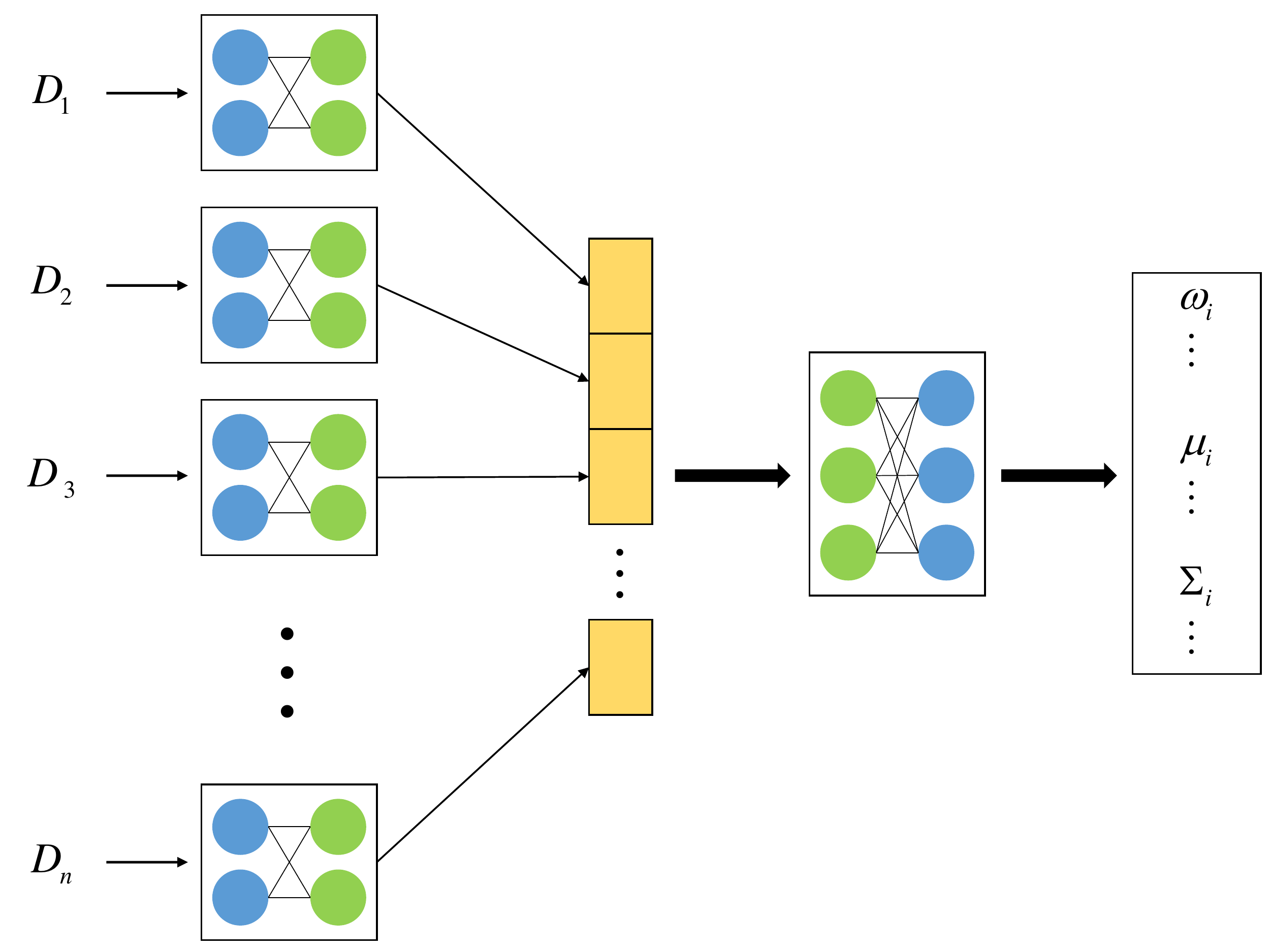}
	\caption{The general structure of a multibranch MDN with Gaussian components. The inputs are multiple data sets $\{D_1, D_2, D_3, ..., D_n\}$ from different experiments, and the outputs are the parameters of the mixture model. The yellow part shows the combination of the outputs of these branches.}\label{fig:mdn_multibranch}
\end{figure}

The MDN model shown in the Figure~\ref{fig:mdn} is for the case of single observational data. But in cosmology we usually need to combine multiple data sets to estimate cosmological parameters. Therefore, following \citet{Wanggj:2020a}, we design a multibranch MDN for a joint constraint on parameters using multiple data sets, as shown in Figure~\ref{fig:mdn_multibranch}. There are multiple branches on the left side of the model that can accept multiple observational data sets as input. Then, the information will be combined in the middle of the model (the yellow part of Figure~\ref{fig:mdn_multibranch}), and finally the MDN outputs the parameters of the mixture model.

The parameters (e.g. $w$ and $b$ in Equation~(\ref{equ:neuron_function})) of an MDN can be learned by minimizing the loss function \citep{Kruse:2020}:
\begin{eqnarray}
\label{equ:loss_mdn}
\mathcal{L} &=& \mathbb{E}\left[-\ln\left(p(\bm\theta|\bm{d})\right)\right] \nonumber \\
&=& \mathbb{E}\left[-\ln\left(\sum_{i=1}^K \omega_i p_i(\bm\theta|\bm{d})\right)\right] ~,
\end{eqnarray}
where the loss is averaged over the minibatch input samples to the network. Specifically, for an MDN with Gaussian components and the case of only one parameter, Equation~(\ref{equ:loss_mdn}) becomes
\begin{align}\label{equ:loss_mdn_gaussian_1}
\mathcal{L} &= \mathbb{E}\left[ -\ln\left(\sum_{i=1}^K \omega_i\cdot\frac{1}{\sqrt{2\pi\sigma^2_i}}e^{-\frac{(\theta-\mu_i)^2}{2\sigma^2_i}}\right)\right] ~.
\end{align}
Because small values in the exponent will give very small values in the logarithm that can lead to numerical underflow, we therefore, use the Log-Sum-Exp trick and carry out the calculations in the logarithmic domain for numerical stability. Thus, Equation~(\ref{equ:loss_mdn_gaussian_1}) can be rewritten as 
\begin{align}
\mathcal{L} &= \mathbb{E}\left[ -\ln\left(\sum_{i=1}^K e^{\left[\ln(\omega_i) + \ln(p_i(\theta|\bm{d}))\right]}\right)\right] ~,
\end{align}
where
\begin{equation}
\ln\left[p_i(\theta|\bm{d})\right] = -\frac{(\theta-\mu_i)^2}{2\sigma_i^2} - \ln(\sigma_i) - \frac{\ln(2\pi)}{2}~.
\end{equation}

For the case with multiple parameters, Equation~(\ref{equ:loss_mdn}) becomes
\begin{eqnarray}
\label{equ:loss_mdn_gaussian_multi}
\mathcal{L} &=&  \mathbb{E}\left[ -\ln\left( \sum_{i=1}^K \omega_i \right.\right. \nonumber \\
&\times& \left.\left. \frac{\exp{\left( -\tfrac{1}{2} (\bm\theta - \bm\mu_i)^\top \bm\Sigma_i^{-1} (\bm\theta - \bm\mu_i) \right)}}{\sqrt{\left( 2\pi \right)^N |\bm\Sigma_i|}} \right)\right] ~.
\end{eqnarray}
Considering the numerical stability, this equation can be written as
\begin{align}
\mathcal{L} &= \mathbb{E}\left[ -\ln\left(\sum_{i=1}^K e^{\left[\ln(\omega_i) + \ln(p_i(\bm\theta|\bm{d}))\right]}\right)\right] ~,
\end{align}
where 
\begin{eqnarray}
\label{equ:pdf_of_gaussian_multi_log}
\ln[p_i(\bm\theta|\bm{d})] &=& -\frac{1}{2} (\bm\theta - \bm\mu_i)^\top \bm\Sigma_i^{-1} (\bm\theta - \bm\mu_i) \nonumber \\
&+& \ln\left(|\bm\Sigma_i^{-1}|^{\frac{1}{2}}\right) - \ln\left(\sqrt{(2\pi)^N}\right).
\end{eqnarray}
Note that for a valid multivariate Gaussian distribution, $\bm\Sigma_i$ and $\bm\Sigma^{-1}_i$ here must be positive-definite matrices. Therefore, the precision matrix $\bm\Sigma^{-1}_i$ can be characterized by its upper Cholesky factor $\bm{U}_i$: 
\begin{align}
\bm\Sigma^{-1}_i = \bm{U}^\top_i\bm{U}_i~,
\end{align}
where $\bm{U}_i$ is an upper triangular matrix with strictly positive diagonal entries. There are $N(N+1)/2$ nonzero entries for $\bm{U}_i$, which is much less than that of $\bm\Sigma^{-1}_i$ ($N^2$ entries). Thus, if the MDN learns $\bm\Sigma^{-1}_i$ instead of $\bm{U}_i$, there will be more neurons in the network, which will increase the training time. Also, the output of the network may also make $\bm\Sigma^{-1}_i$ a non-positive-definite matrix. Therefore, in order to make the convergence of the network more stable, the MDN is actually learns $\bm{U}_i$. Outputs of the network can be either positive or negative numbers. Thus, to ensure the diagonal entries of $\bm{U}_i$ are positive, we use a Softplus function to enforce the positiveness of the diagonal entries:
\begin{align}
(\bm{U}_i)_{jk} &=
\begin{cases}
\text{Softplus}(\widetilde{\bm{U}}_i)_{jk}~, & \text{if } j=k \\
(\widetilde{\bm{U}}_i)_{jk}~, & \text{otherwise}
\end{cases}
\end{align}
where $\widetilde{\bm{U}}_i$ is the output of the network, and 
\begin{equation}\label{equ:softplus}
\text{Softplus}(x) = \frac{1}{\beta}\ln(1 + e^{\beta x})~,
\end{equation}
where $\beta$ is a parameter and we set $\beta=1$ throughout the paper. This offers an efficient way to calculate the matrix in Equation~(\ref{equ:pdf_of_gaussian_multi_log}):
\begin{align}
\nonumber|\bm\Sigma_i^{-1}| &= |\bm{U}^\top_i\bm{U}_i| \\
\nonumber&= |\bm{U}^\top_i|\cdot|\bm{U}_i| \\
&= \left(\prod_{j=1}^N\text{diag}(\bm{U}_i)_j \right)^2~.
\end{align}
Therefore,
\begin{equation}
\ln\left(|\bm\Sigma_i^{-1}|^{\frac{1}{2}}\right) = \sum_{j=1}^N\ln\left(\text{diag}(\bm{U}_i)_j\right)~.
\end{equation}
In addition,
\begin{eqnarray}
(\bm\theta - \bm\mu_i)^\top \bm\Sigma_i^{-1} (\bm\theta - \bm\mu_i) &=& (\bm\theta - \bm\mu_i)^\top \bm{U}^\top_i\cdot\bm{U}_i (\bm\theta - \bm\mu_i) \nonumber \\
&=& \left[\bm{U}_i (\bm\theta - \bm\mu_i)\right]^\top\cdot\left[\bm{U}_i (\bm\theta - \bm\mu_i)\right] \nonumber \\
&=& \norm{\bm{U}_i (\bm\theta - \bm\mu_i)}^2_2.
\end{eqnarray}
Therefore, Equation~(\ref{equ:pdf_of_gaussian_multi_log}) can be finally written as
\begin{align}
\nonumber \ln[p_i(\bm\theta|\bm{d})] &= -\frac{1}{2}\norm{\bm{U}_i (\bm\theta - \bm\mu_i)}^2_2 + \sum_{j=1}^N\ln\left(\text{diag}(\bm{U}_i)_j\right) \\
&\quad - \ln\left(\sqrt{(2\pi)^N}\right)~.
\end{align}

\subsection{Training Strategy}\label{sec:training_strategy}
For parameter estimation with the MDN, we will first train the MDN with simulated data and then estimate parameters with the observational data. In this section, we describe the processes of training and parameter estimation.

\subsubsection{Training Set}\label{sec:training_set}

The basic principle that the MDN can be used for parameter estimation lies in the fact that it can learn a mapping from the input data space to the output parameter space of the mixture model. Note that the parameters of the mixture model are functions of the cosmological parameters to be estimated. Therefore, the parameter space of the cosmological model in the training set should be large enough to cover the posterior distribution of the cosmological parameters. Following \citet{Wanggj:2020a}, we set the range of parameters in the training set to $[P-5\sigma_{p}, P+5\sigma_{p}]$, where $P$ is the best-fitting value of the posterior distribution of cosmological parameters and $\sigma_p$ is the corresponding $1\sigma$ error. Note that the best-fitting value here refers to the mode of the posterior distribution. In this parameter space, the cosmological parameters in the training set are simulated according to the uniform distribution.

From the illustrations of Sections \ref{sec:gaussian_mixture_model} and \ref{sec:mdn}, one can see that the error of the cosmological parameters should be calculated by the standard deviation $\sigma$ or covariance matrix $\bm\Sigma$ of the mixture model, which should be determined by the covariance matrix of the observational measurement. This means that the MDN should know what the error looks like. Therefore, inspired by \citet{Wanggj:2020a}, we add noise to the training set based on the covariance matrix of the observational measurement. Specifically, Gaussian noise $\mathcal{N}(0, \bm\Sigma)$ is generated and added to the samples of the training set at each epoch of the training process. Note that the noise level is equal to the observational errors, which is different from \citet{Wanggj:2020a} where multiple levels of noise can be added to the training set. For more details, we refer interested readers to~\citet{Wanggj:2020a}.

The orders of magnitude of the cosmological parameters may be quite different, which will have an effect on the performance of the network. To address this, data preprocessing techniques should be used before feeding the data to the MDN. Following \citet{Wanggj:2020a}, we conduct two steps to normalize the data. First, we divide the cosmological parameters in the training set by their estimated values (here the mean of the parameters in the training set is taken as the estimated value) to ensure that they all become of order of unity; then, we normalize the training set using the $z$-score normalization technique \citep{Kreyszig:2011}:
\begin{eqnarray}
z = \frac{x-\mu}{\sigma}~,
\end{eqnarray}
where $\mu$ and $\sigma$ are the mean and standard deviation of the measurement $\bm{d}$ or the corresponding cosmological parameters $\bm\theta$. This data preprocessing method can reduce the influence of differences between parameters on the result, thus making the MDN a general method that can be applied to any cosmological parameters.

\subsubsection{Training and Parameter Estimation}\label{sec:training_parameter_estimation}
Here we illustrate how to train an MDN and estimate parameters using a well-trained MDN model. The key steps of the training process are as follows:

\begin{enumerate}
	\item Provide a cosmological model and the initial conditions of the corresponding cosmological parameters.
	\item\label{list:step2} Simulate the training set using the method illustrated in Section \ref{sec:training_set}.
	\item Add noise and conduct data preprocessing on the training set using the method illustrated in Section \ref{sec:training_set}.
	\item Feed the training set to the MDN and train the model.
	\item Feed the observational measurement to the well-trained MDN model to obtain the parameters of the mixture model.
	\item Obtain an ANN chain by generating samples using the mixture model according to Equations~(\ref{equ:pdf_of_gaussian_1}) and (\ref{equ:pdf_of_gaussian_multi}).
	\item\label{list:step7} Calculate the best-fitting values and errors of the cosmological parameters by using the ANN chain. Then, the parameter space to be learned will be updated according to the estimated parameters.
	\item By repeating steps~\ref{list:step2}-\ref{list:step7}, one can obtain a stable ANN chain for parameter estimation.
\end{enumerate}

Note that the initial conditions of cosmological parameters set in step 1 are general ranges of the parameters. This means that the true parameters can be located outside of these ranges. Therefore, the parameter space should be updated in step 7 to continue training.

\section{\bf Application to SN-Ia}\label{sec:apply_to_panthon}
In this section, we apply the MDN method to the SN-Ia data set to estimate parameters of the $w$CDM cosmology.

\subsection{SN-Ia data}\label{sec:pantheon_data}

To test the capability of the MDN for estimating parameters from one data set, we constrain $w$ and $\Omega_{\rm m}$ of the $w$CDM model using the Pantheon SN-Ia \citep{Scolnic:2018}, which contain 1048 data points for the redshift range [0.01, 2.26]. The distance modulus of the Pantheon SN-Ia can be rewritten as
\begin{equation}
\mu=m_{B,{\rm corr}}^* - M_{B} ~,
\end{equation}
where $m_{B,{\rm corr}}^* = m_{B}^*+\alpha\times x_1-\beta\times c + \Delta_B$ is the corrected apparent magnitude, and $M_B$ is the $B$-band absolute magnitude~\citep{Scolnic:2018}. The variable $\alpha$ is the coefﬁcient of the relation between luminosity and stretch, $\beta$ is the coefﬁcient of the relation between luminosity and color, and $\Delta_B$ is a distance correction based on predicted biases from simulations.

The measurement of the Pantheon SN-Ia is the corrected apparent magnitudes. Therefore, the input of the MDN is $m_{B,{\rm corr}}^*$ for the observational SN-Ia, and $\mu + M_B$ for the simulated data generated by the $w$CDM model. Specifically, we have the distance modulus
\begin{align}\label{equ:mu_wCDM}
\mu = 5\log_{10}\left(D_{\rm L}(z)/{\rm Mpc}\right) + 25 ~,
\end{align}
and the luminosity distance
\begin{equation}\label{equ:dl_wCDM}
D_{\rm L}(z)=\frac{c(1+z)}{H_0}\int_0^z\frac{{\rm d}z^\prime}{E(z^\prime)}~,
\end{equation}
where $c$ is the speed of light, and $H_0$ is the Hubble constant. The $E(z)$ function describes the evolution of the Hubble parameter:
\begin{equation}
E(z) = \sqrt{\Omega_{\rm m}(1+z)^3 + (1-\Omega_{\rm m})(1+z)^{3(1+w)}}~, \label{eq:Ez}
\end{equation}
where $\Omega_{\rm m}$ is the matter density parameter and $w$ is the equation of state of dark energy. Equation~(\ref{eq:Ez}) assumes the spatial flatness. Then we have
\begin{align}\label{equ:mu_wCDM_plus_MB}
\mu + M_B &= 5\log_{10}\frac{D_{\rm L}(z)}{\rm Mpc} + M_B + 25~.
\end{align}
As we already know, the absolute magnitude of SN-Ia $M_B$ is strongly degenerated with the Hubble constant $H_0$. Therefore, we combine $M_B$ and $H_0$ into a new parameter and constrain it with the cosmological parameters simultaneously. To do this, we define a dimensionless luminosity distance
\begin{equation}
\widetilde{D}_{\rm L}(z) = (1+z)\int_0^z\frac{{\rm d}z^\prime}{E(z^\prime)}~.
\end{equation}
Then Equation~(\ref{equ:mu_wCDM_plus_MB}) can be rewritten as 
\begin{align}
\nonumber\mu + M_B &= 5\log_{10}\frac{(c/H_{0})\cdot \widetilde{D}_{\rm L}(z)}{\rm Mpc} + M_B + 25\\
&= 5\log_{10} \widetilde{D}_{\rm L}(z) + \mu_c~,
\end{align}
where $\mu_c \equiv 5\log_{10}\left(c/H_0/ {\rm Mpc}\right) + M_B + 25$ is a new nuisance parameter to be estimated with the cosmological parameters simultaneously.

\begin{table}
	\centering
	\caption{Best-fitting values of the $w$CDM parameters using the Pantheon SN-Ia data without considering the systematic covariance matrix. The error is $1\sigma$ C.L.}\label{tab:params_pantheon}
	\begin{tabular}{c|c|c}
		\hline\hline
		& \multicolumn{2}{c}{Methods} \\
		\cline{2-3}
		Parameters & MCMC & MDN \\
		\hline
		$w$ & $-1.235\pm0.141$ & $-1.230\pm0.139$ \\
		$\Omega_{\rm m}$ & $0.356\pm0.033$ & $0.351\pm0.034$ \\
		\hline\hline
	\end{tabular}
\end{table}
\begin{figure}
	\centering
	\includegraphics[width=0.45\textwidth]{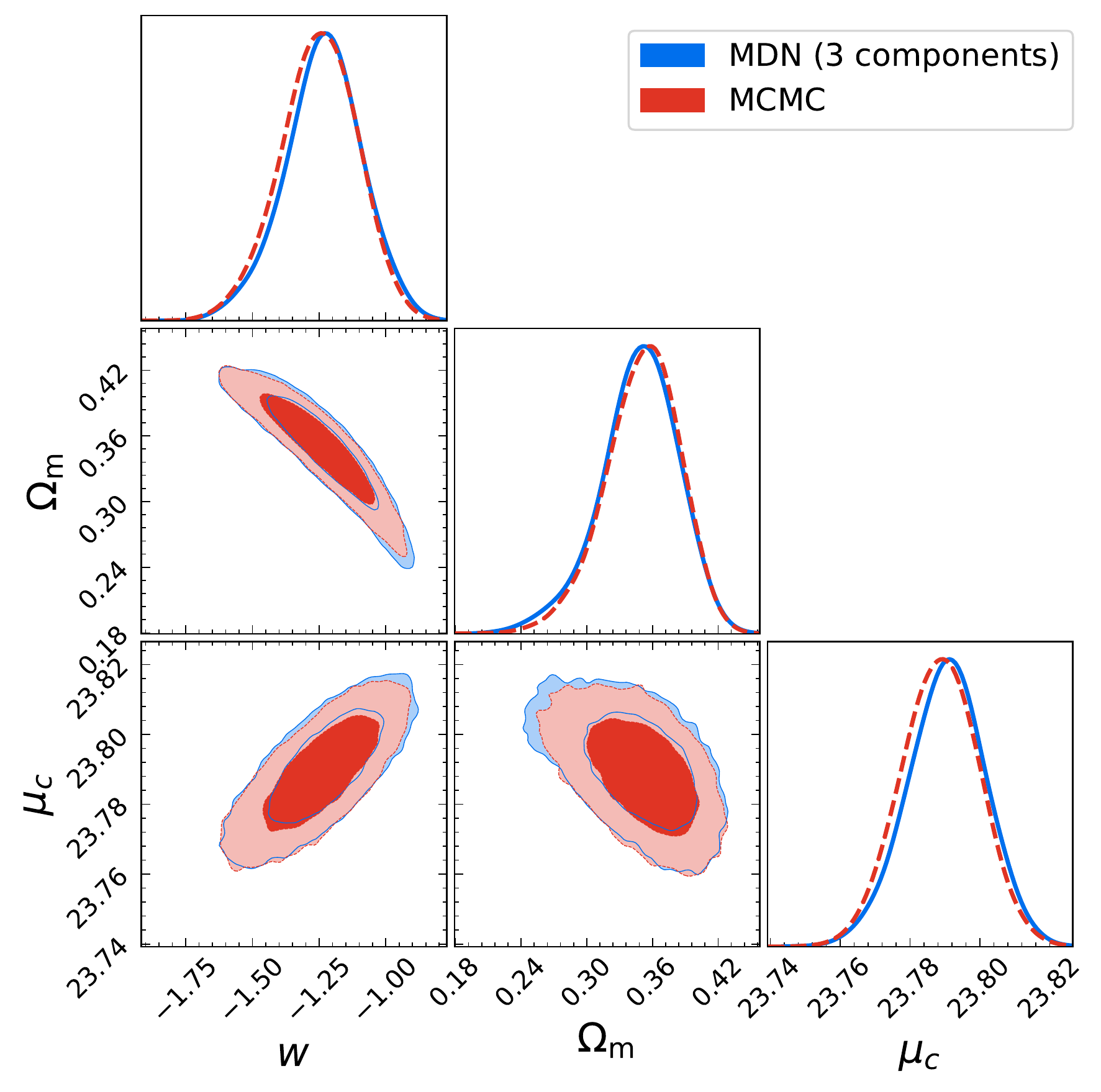}
	\caption{One-dimensional and two-dimensional marginalized distributions with 1$\sigma$ and 2$\sigma$ contours of $w$, $\Omega_{\rm m}$, and $\mu_c$ constrained from Pantheon SN-Ia, with no systematic uncertainty. Three Gaussian components are used here for the MDN method.}\label{fig:contour_pantheon}
\end{figure}
\begin{figure}
	\centering
	\includegraphics[width=0.45\textwidth]{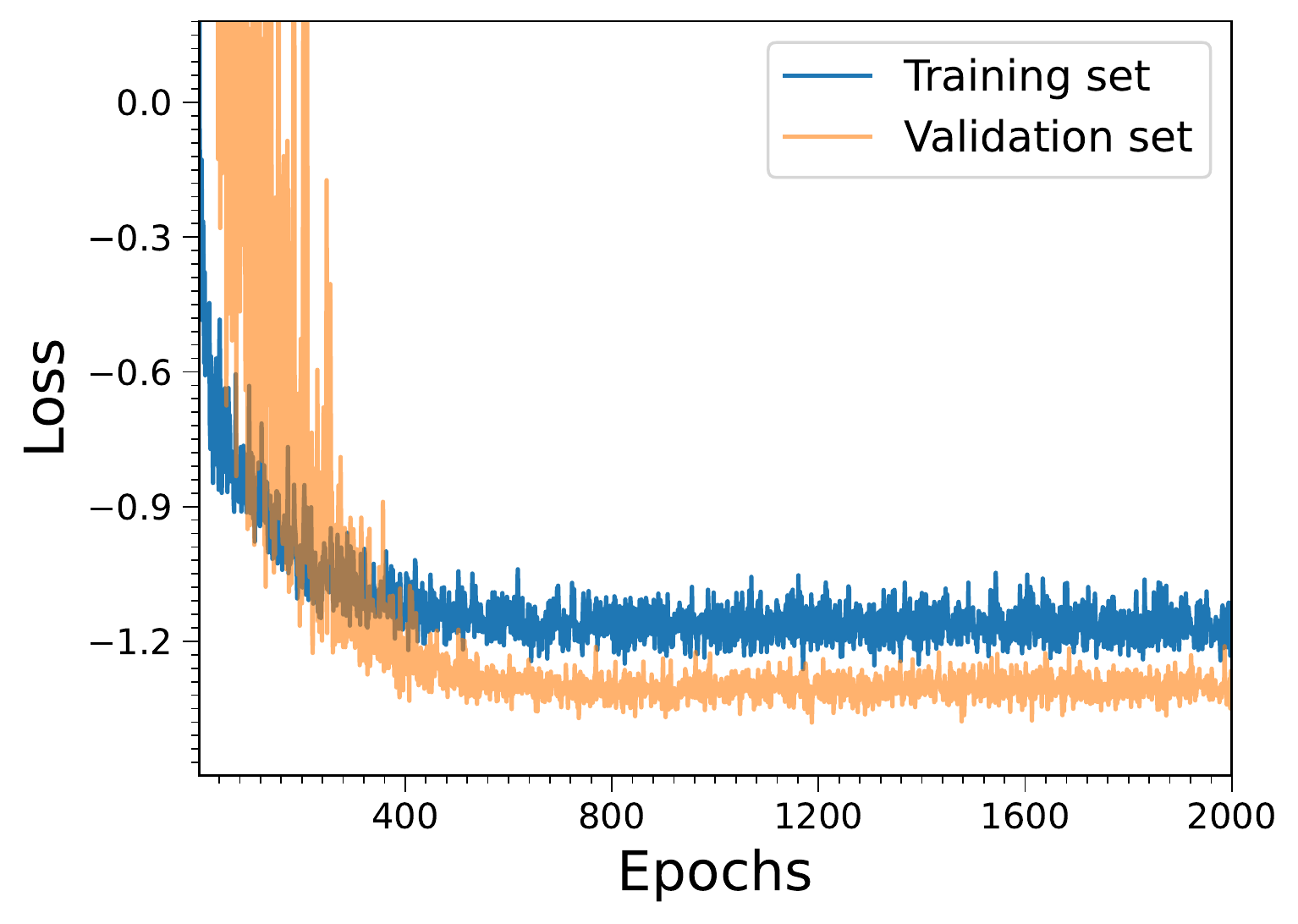}
	\caption{Losses of the training set and validation set. There are 3000 samples in the training set, and 500 samples in the validation set.}\label{fig:loss_pantheon}
\end{figure}

\subsection{Result}\label{sec:pantheon_fitting_result}

We first constrain $w$, $\Omega_{\rm m}$, and the nuisance parameter $\mu_c$ simultaneously using the Pantheon SN-Ia data without considering systematic uncertainties. Here two methods are considered in our analysis: standard MCMC and the MDN. For the MCMC method, we use the public package {\it emcee} \citep{Foreman-Mackey:2013} for parameter estimation, which is a Python module that achieves the MCMC method. During the constraining procedures, an MCMC chain with 100,000 steps is generated after burn-in. The best-fitting values with 1$\sigma$ errors of the parameters are calculated from the MCMC chains with {\it coplot},\footnote{\url{https://github.com/Guo-Jian-Wang/coplot}} as shown in Table \ref{tab:params_pantheon}. We also draw the corresponding one-dimensional and two-dimensional contours in Figure~\ref{fig:contour_pantheon}. For the MDN method, we consider three Gaussian components and using the setting illustrated in Sections \ref{sec:ann}, \ref{sec:mdn}, and \ref{sec:training_strategy}. There are 3000 samples in the training set and 500 samples in the validation set. After training, we feed the corrected apparent magnitudes $m_{B,{\rm corr}}^*$ to the well-trained MDN model to obtain the parameters of the mixture model, and then get an ANN chain of the cosmological parameters. The results are shown in Table \ref{tab:params_pantheon} and Figure~\ref{fig:contour_pantheon}. Obviously, the results of the MDN method are almost the same as those of the MCMC method. 

We show the losses of the training and validation sets in Figure~\ref{fig:loss_pantheon}. The loss of the training set is large at the beginning of the training process, and gradually decreases and stabilizes with the epoch. For the validation set, the trend of loss is similar to that of the training set. After 400 epochs, the loss of the validation set is smaller than that of the training set. Obviously, there is no overfitting for the MDN. The reason why the loss of the validation set is smaller than that of the training set is that multiple sets of noise are added to each sample, which means that multiple sets of noise will be generated and added to a sample to ensure that the MDN knows that measurement may differ due to the presence of noise. Our analysis shows that the adding of multiple sets of noise to measurement is beneficial to the stability of the MDN and will make the final estimated cosmological parameters more accurate, especially for the multibranch MDN.

\begin{figure}
	\centering
	\includegraphics[width=0.45\textwidth]{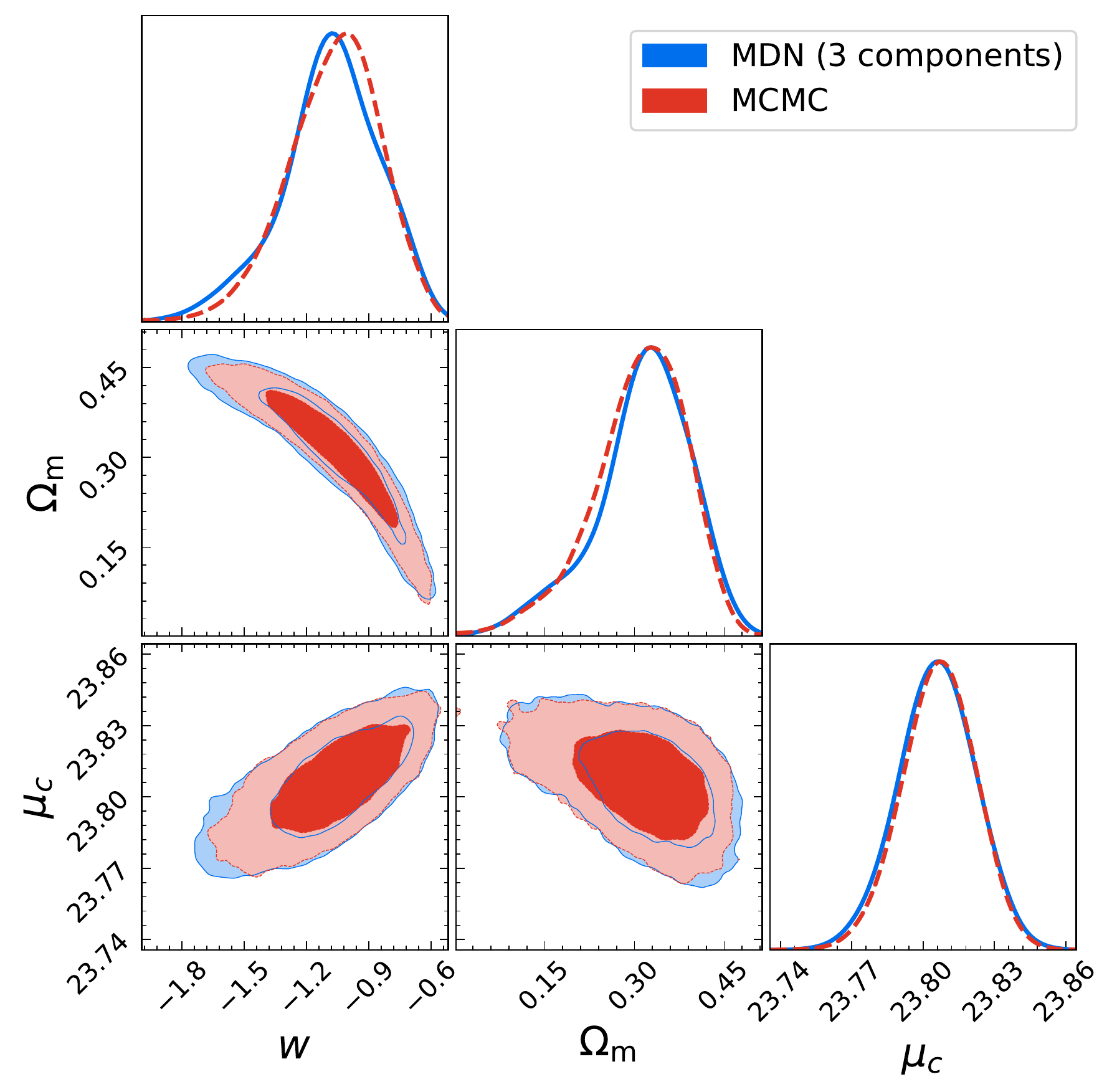}
	\caption{The same as Figure~\ref{fig:contour_pantheon}, but including systematic uncertainties.}\label{fig:contour_pantheon_cov}
\end{figure}

Furthermore, we consider the systematic uncertainties in our analysis to test the capability of the MDN for dealing with observations with covariance. We use the systematic covariance matrix $\bm C_{\rm sys}$ \citep{Scolnic:2018} to generate noise then add it to the training set (see Section \ref{sec:training_set}). We then constrain the parameters with the MCMC and MDN (with three Gaussian components) methods. The results are shown in Table \ref{tab:params_pantheon_cov} and Figure~\ref{fig:contour_pantheon_cov}. The difference between the MDN results and the MCMC results can then be calculated using
\begin{equation}
\Delta P = \frac{|P_{\rm MCMC} - P_{\rm MDN}|}{\sigma} ,
\end{equation}
where $\sigma=\sqrt{\sigma^2_{\rm MCMC} + \sigma^2_{\rm MDN}}$, and $P_{\rm MCMC}$ and $P_{\rm MDN}$ are the best-fitting parameters for the MCMC and MDN methods, respectively. Specifically, the differences for the parameters are $0.174\sigma$, $0.015\sigma$, and $0.076\sigma$, respectively, which are slightly larger for the parameter $w$.

\begin{table}
	\centering
	\caption{The same as Table \ref{tab:params_pantheon}, but including systematic uncertainties.}\label{tab:params_pantheon_cov}
	\begin{tabular}{c|c|c}
		\hline\hline
		& \multicolumn{2}{c}{Methods} \\
		\cline{2-3}
		Parameters & MCMC & MDN \\
		\hline
		$w$         & $-1.011\pm0.216$    & $-1.060\pm0.229$    \\
		$\Omega_{\rm m}$ & $0.327\pm0.074$ & $0.328\pm0.076$ \\
		\hline\hline
	\end{tabular}
\end{table}

\begin{figure}
	\centering
	\includegraphics[width=0.45\textwidth]{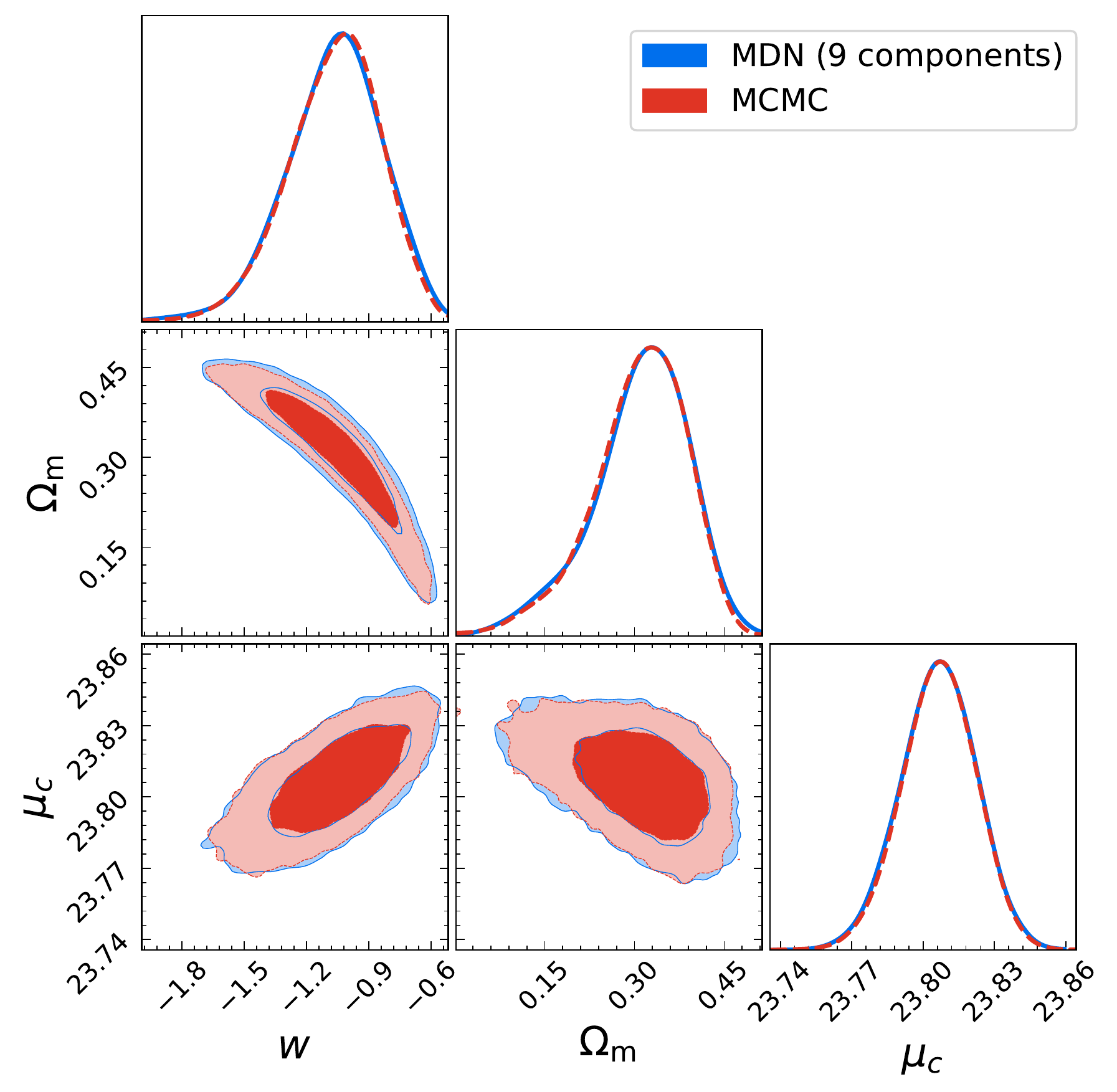}
	\caption{The same as Figure~\ref{fig:contour_pantheon_cov}, but nine Gaussian components are used here.}\label{fig:contour_pantheon_cov_comp9}
\end{figure}

We can see that the shapes of both the one-dimensional and two-dimensional contours based on the MDN method are slightly different from those of the MCMC method. The reason is that the joint probability distribution of $w$ and $\Omega_{\rm m}$ deviates from the multivariate Gaussian distribution, which makes it more difficult to model their joint probability distribution. We take two approaches to solve this problem. One is to increase the number of components ($p_{i}$ function in Equation~(\ref{equ:pdf_of_mixture_model})), which diversifies the function forms that the MDN can fit. Specifically, we use nine Gaussian components and reestimate the cosmological parameters with the MDN, then plot the one-dimensional and two-dimensional contours in Figure~\ref{fig:contour_pantheon_cov_comp9}. One can see that the MDN result converges to the MCMC result. Quantitatively, the differences between the MDN results and the MCMC results are $0.051\sigma$, $0.046\sigma$, and $0.025\sigma$. These deviations are smaller than those of using three components, especially for parameter $w$. Another way is to more efficiently sample the parameter space. As we illustrated in Section \ref{sec:training_set}, samples are generated uniformly in the $\pm 5\sigma_p$ range of the posterior distribution, which might not be the most efficient way. We will explore different sampling methods in future works.

\begin{table}
	\centering
	\caption{1$\sigma$ Constraints on parameters of the $\Lambda$CDM model using the {\it Planck}-2015 CMB temperature and polarization power spectra and Pantheon SN-Ia.}\label{tab:params_planck_pantheon}
	\begin{tabular}{c|c|c}
		\hline\hline
		& \multicolumn{2}{c}{Methods} \\
		\cline{2-3}
		Parameters & MCMC & MDN \\
		\hline
		$H_0$               & $67.701\pm0.633$    & $67.698\pm0.619$ \\
		$\Omega_{\rm b}h^2$ & $0.02231\pm0.00015$ & $0.02229\pm0.00015$ \\
		$\Omega_{\rm c}h^2$ & $0.11866\pm0.00141$ & $0.11867\pm0.00138$ \\
		$\tau$		        & $0.06589\pm0.01346$ & $0.06170\pm0.01412$ \\
		$10^9A_{\rm s}$     & $2.13366\pm0.05636$ & $2.11930\pm0.05908$ \\
		$n_{\rm s}$         & $0.96817\pm0.00398$ & $0.96787\pm0.00382$ \\
		$M_B$		        & $-19.418\pm0.017$   & $-19.418\pm0.017$ \\
		\hline\hline
	\end{tabular}
\end{table}
\begin{figure*}
	\centering
	\includegraphics[width=0.96\textwidth]{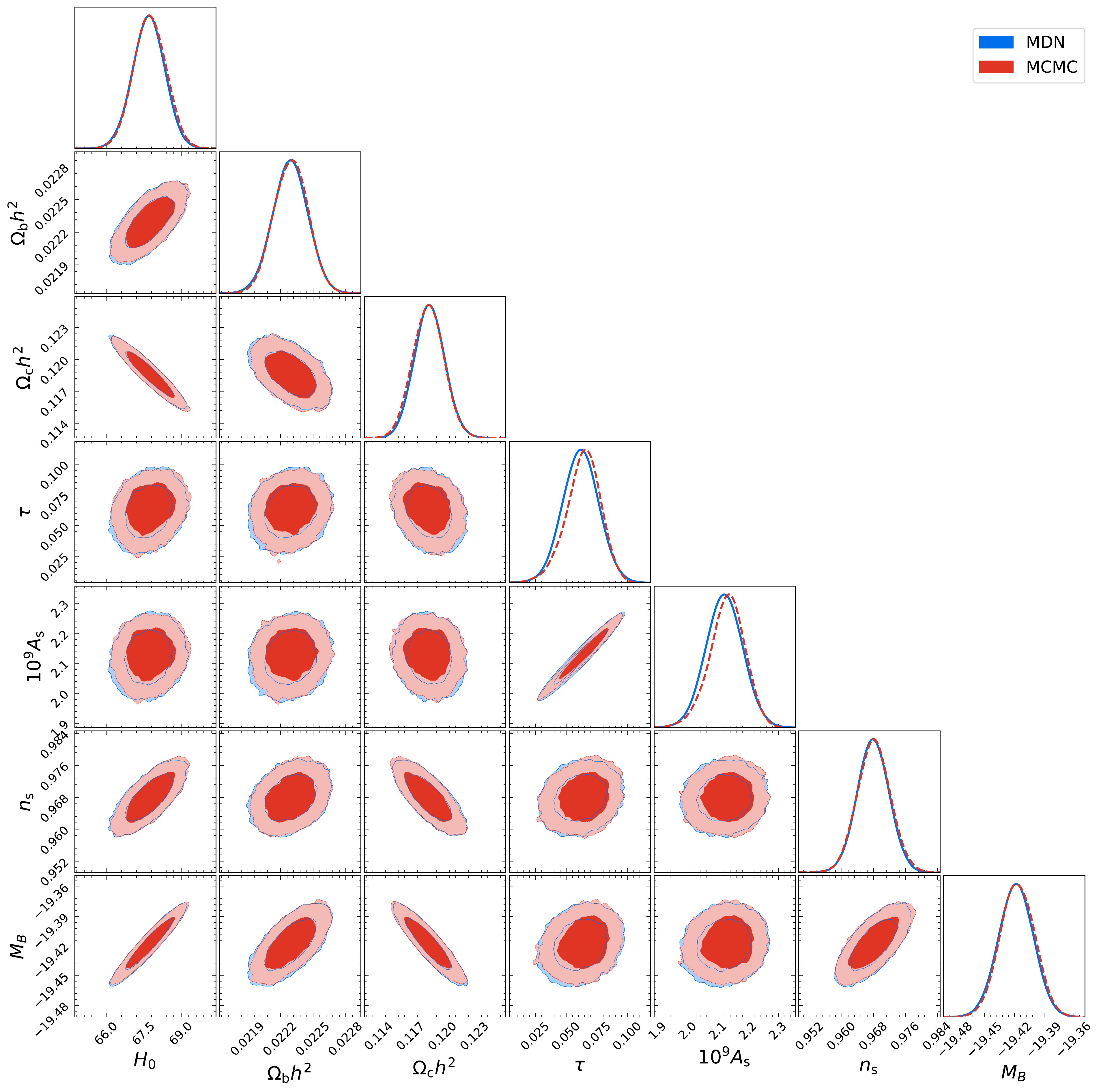}
	\caption{One-dimensional and two-dimensional marginalized distributions with 1$\sigma$ and 2$\sigma$ contours of $H_0$, $\Omega_{\rm b}h^2$, $\Omega_{\rm c}h^2$, $\tau$, $A_{\rm s}$, $n_{\rm s}$, and $\sum m_\nu$ constrained from {\it Planck}-2015 CMB temperature and polarization power spectra and Pantheon SN-Ia. Three Gaussian components are used here for the MDN method.}\label{fig:contour_planck_pantheon}
\end{figure*}

\begin{figure*}
	\centering
	\includegraphics[width=0.45\textwidth]{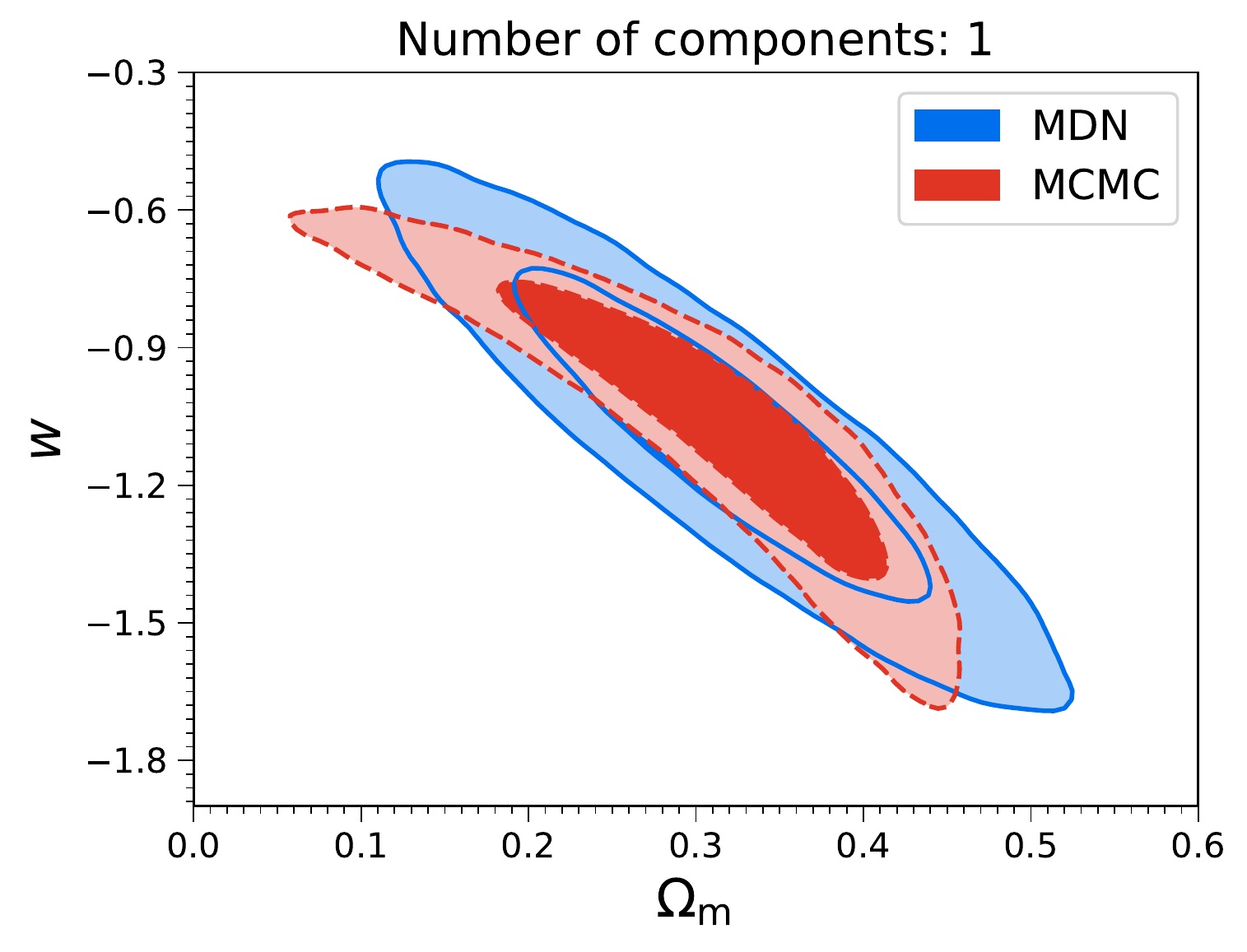}
	\includegraphics[width=0.45\textwidth]{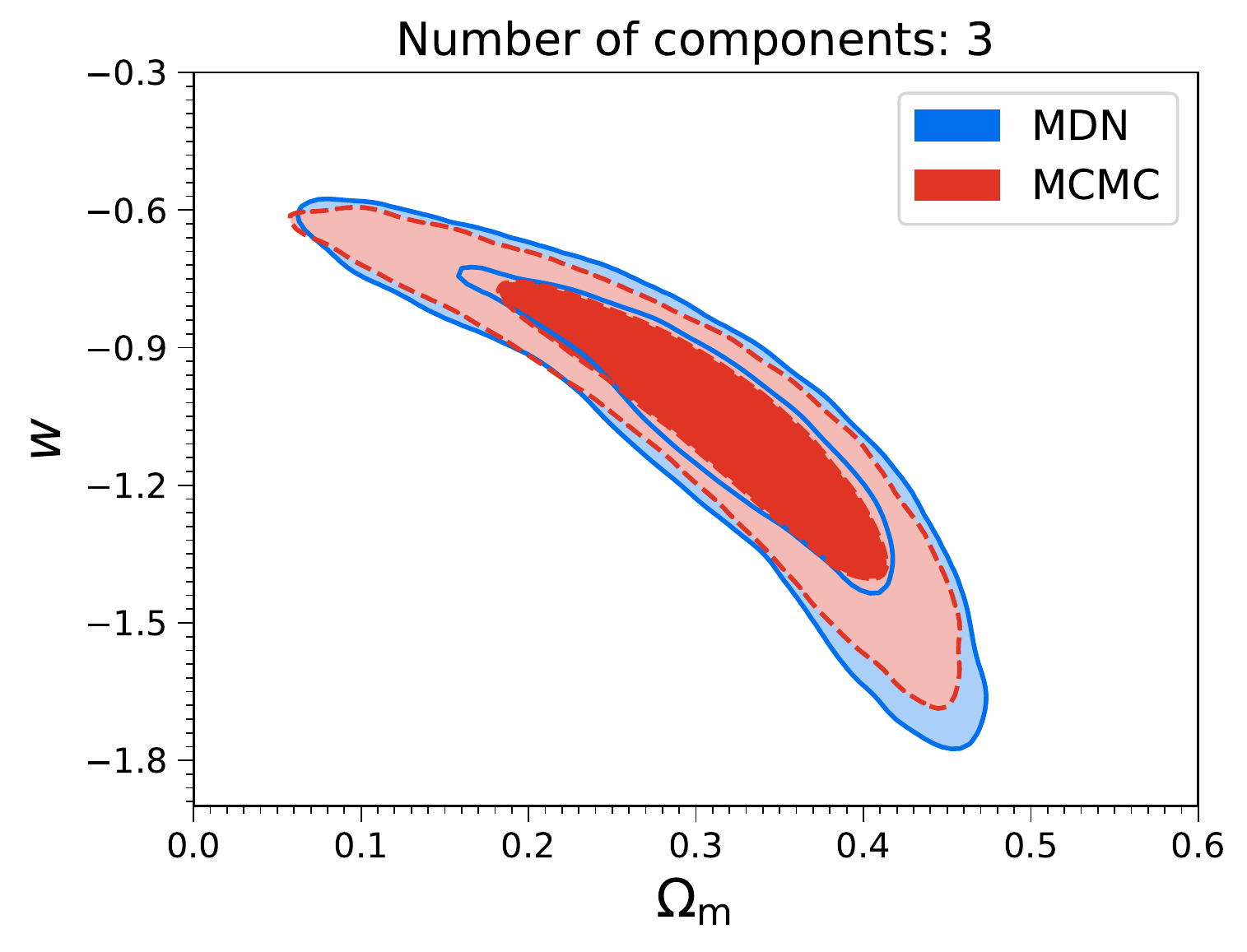}
	\includegraphics[width=0.45\textwidth]{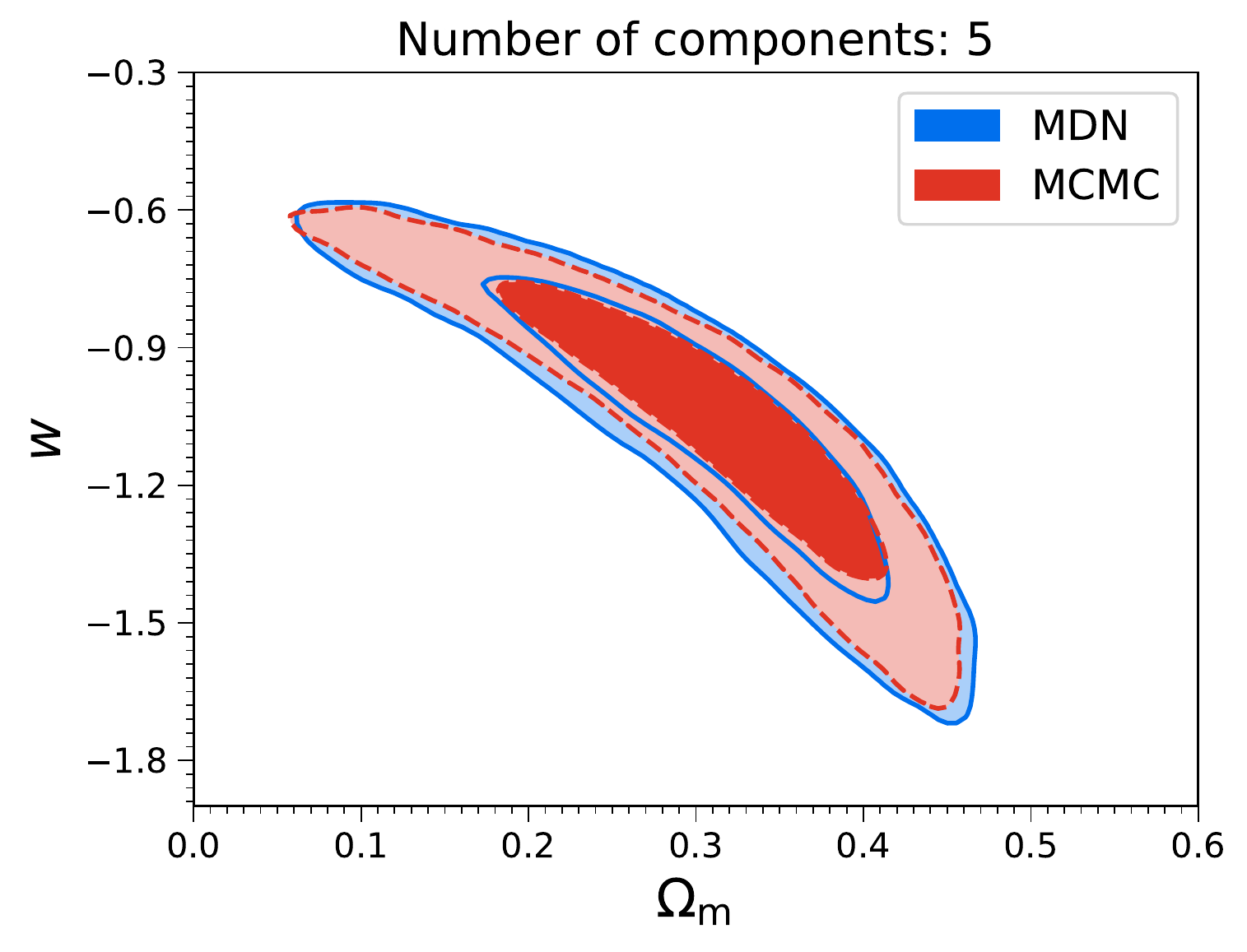}
	\includegraphics[width=0.45\textwidth]{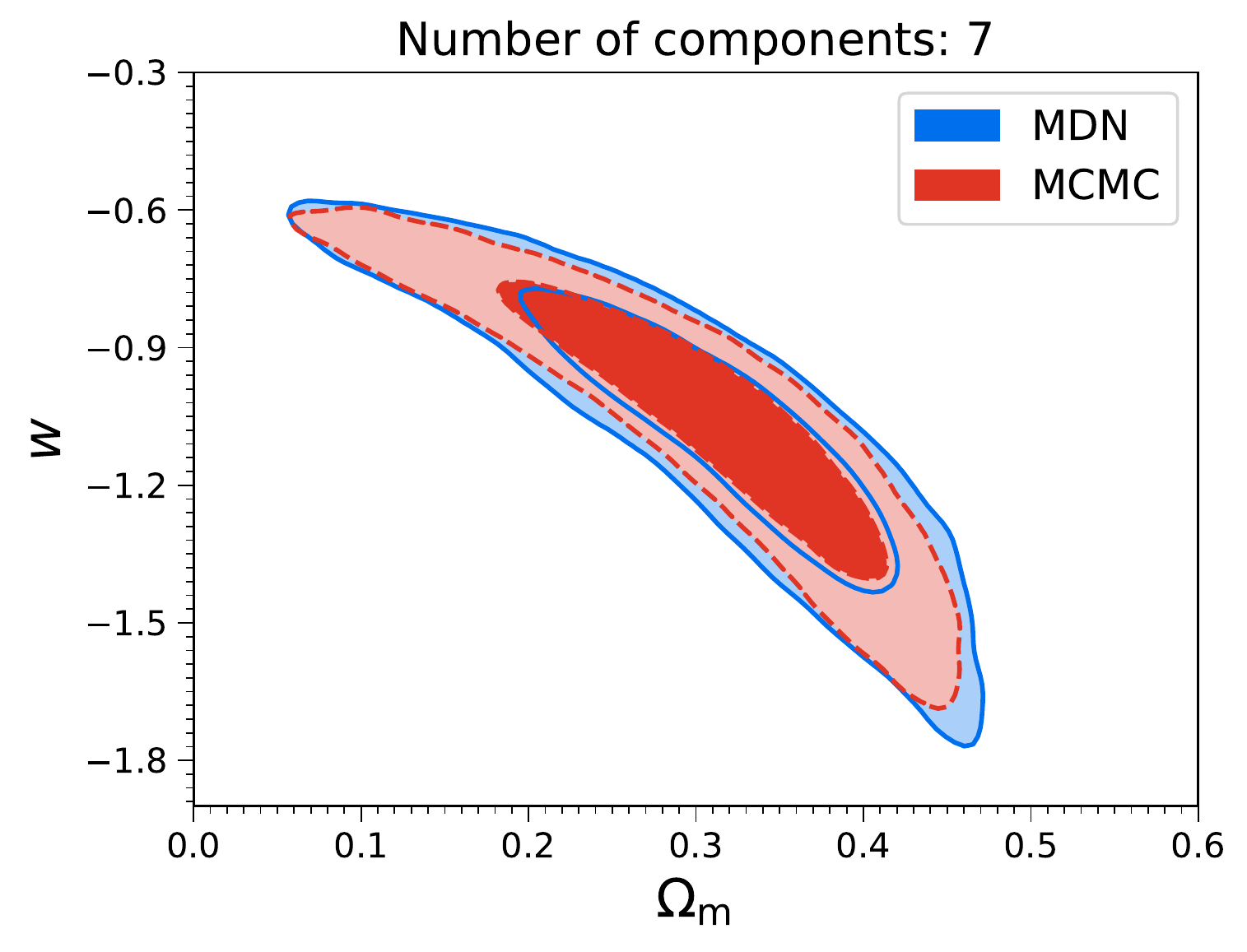}
	\caption{Two-dimensional marginalized distributions with 1$\sigma$ and 2$\sigma$ contours of $w$ and $\Omega_{\rm m}$ constrained from Pantheon SN-Ia when considering the systematic uncertainties.}\label{fig:effect_of_component}
\end{figure*}

\section{\bf Joint Constraint on Parameters}\label{sec:joint_constraint_on_parameters}

Cosmological research often combines multiple data sets to constrain cosmological parameters. So, in this section, we estimate parameters using the multibranch MDN. Our purpose is to verify the feasibility of this method. Therefore, for simplicity of our analysis, we will only consider the use of {\it Planck}-2015 temperature and polarization power spectra {COM\_PowerSpect\_CMB\_R2.02.fits}\footnote{\url{http://pla.esac.esa.int/pla/\#cosmology}} and Pantheon SN-Ia data to constrain parameters. These parameters include the $B$-band absolute magnitude $M_B$ (nuisance parameter in SN-Ia) and six parameters in the standard $\Lambda$CDM model, i.e. the Hubble constant $H_0$, the baryon density $\Omega_{\rm b}h^2$, the cold dark matter density $\Omega_{\rm c}h^2$, the optical depth $\tau$, and the amplitude and spectral index of primordial curvature perturbations ($A_{\rm s}$, $n_{\rm s}$). There are four branches in the multibranch MDN model, and each of them accepts one measurement, namely the TT, EE, and TE power spectra of the CMB, and the corrected apparent magnitudes $m_{B,{\rm corr}}^*$ of the Pantheon SN-Ia. We calculate the power spectra of the CMB with the public code {\sc CAMB}\footnote{\url{https://github.com/cmbant/CAMB}}~\citep{camb}.

We first constrain these parameters using the MCMC method, and an MCMC chain with 100,000 steps is generated. Then the best-fitting values with the $1\sigma$ errors of these parameters are calculated using the MCMC chain, as shown in Table \ref{tab:params_planck_pantheon}. The corresponding one-dimensional and two-dimensional marginalized distributions with $1\sigma$ and $2\sigma$ contours are shown in Figure~\ref{fig:contour_planck_pantheon}.

Then, following the same procedure in Section \ref{sec:apply_to_panthon}, we constrain the parameters using the MDN method. In the training process, 3000 samples are used to train the MDN model, and 500 samples are used as a validation set. After training of the MDN model, we feed the observational TT, EE, and TE power spectra of the {\it Planck} CMB, as well as the corrected apparent magnitudes $m_{B,{\rm corr}}^*$ of the Pantheon SN-Ia, to the MDN model to obtain the parameters of the mixture model. Finally, we generate samples using the mixture model to obtain an ANN chain of the cosmological parameters. The best-fitting values and $1\sigma$ errors calculated using the chain are shown in Table \ref{tab:params_planck_pantheon}, and the corresponding one-dimensional and two-dimensional contours are shown in Figure~\ref{fig:contour_planck_pantheon}. We can see that both the best-fitting values and the $1\sigma$ errors are almost the same as those of the MCMC method. The deviations of the parameters between the two methods are $0.003\sigma$, $0.078\sigma$, $0.007\sigma$, $0.215\sigma$, $0.176\sigma$, $0.055\sigma$, and $0.006\sigma$, which are quite small on average. This indicates the MDN method can combine multiple observational data to jointly constrain cosmological parameters, which is generally applicable to all cosmological parameter estimations.

\section{\bf Effect of Hyperparameters}\label{sec:effect_of_hyperparameters}
There are many hyperparameters that can be selected manually in the MDN model, such as the number of components, the number of hidden layers, the activation functions, the number of training samples, and the number of epochs. In this section, we discuss the effect of these hyperparameters on results. All results in this section are evaluated on observational data or simulated observations that can be taken as the test set.

\subsection{The Number of Components}\label{sec:effect_of_components}

As we have shown in Section \ref{sec:pantheon_fitting_result} the number of components of the mixture model may affect the constraints of parameters. Here, we further discuss the effect of the number of components on the result. Specifically, with the same procedure, we estimate $w$ and $\Omega_{\rm m}$ of the $w$CDM model with Pantheon SN-Ia data by considering one, three, five, and seven Gaussian components. The two-dimensional contours of the parameters are shown in Figure~\ref{fig:effect_of_component}.

For the case of one component, the constraint is
\begin{align}
w &= -1.100\pm 0.237, & \Omega_{\rm m} &= 0.314\pm 0.081.
\end{align}
This result converges to the MCMC result (see Table \ref{tab:params_pantheon_cov}), and both the best-fitting values and $1\sigma$ errors are similar to those of the MCMC method. However, the two-dimensional contour (the upper left panel of Figure~\ref{fig:effect_of_component}) is quite different from the contour of the MCMC method. The reason, as we illustrated in Section \ref{sec:pantheon_fitting_result}, is that the joint probability distribution of $w$ and $\Omega_{\rm m}$ deviates significantly from the multivariate Gaussian distribution.

For the cases with three, five, and seven components, the results all converge to the MCMC result, and all the shapes of the two-dimensional contours are similar to the MCMC result. This suggests that for $w$ and $\Omega_{\rm m}$ of the $w$CDM model, three Gaussian components can characterize the joint probability distribution. Therefore, we recommend using multiple components, such as three or more, when estimating parameters of other models. However, it is not recommended to use a very large number of components (e.g. more than 10), especially for multiple-parameters cases, because the MDN will be unstable and thus it will become difficult to constrain these components.

\begin{figure*}
	\centering
	\includegraphics[width=0.45\textwidth]{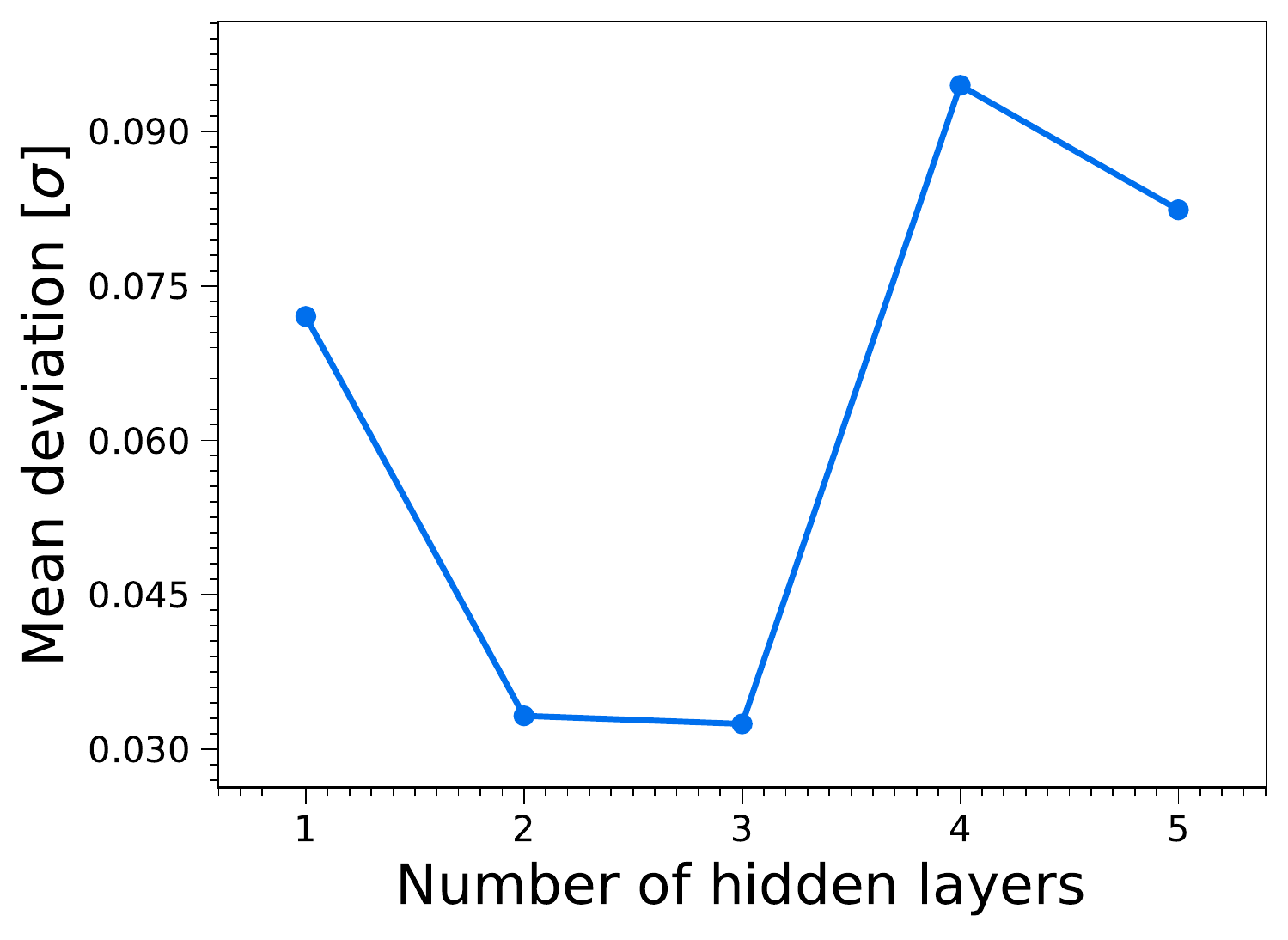}
	\includegraphics[width=0.45\textwidth]{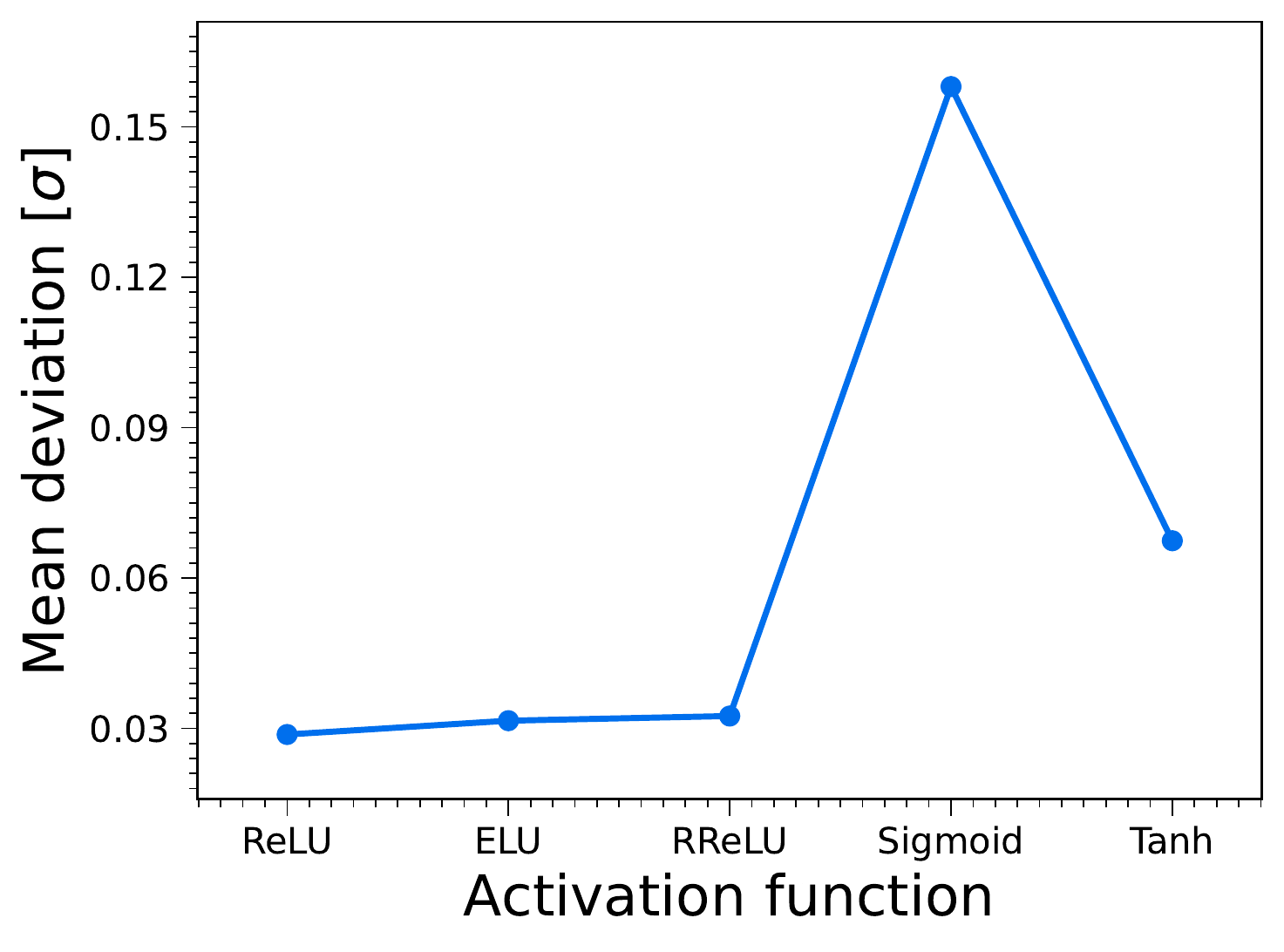}
	\includegraphics[width=0.448\textwidth]{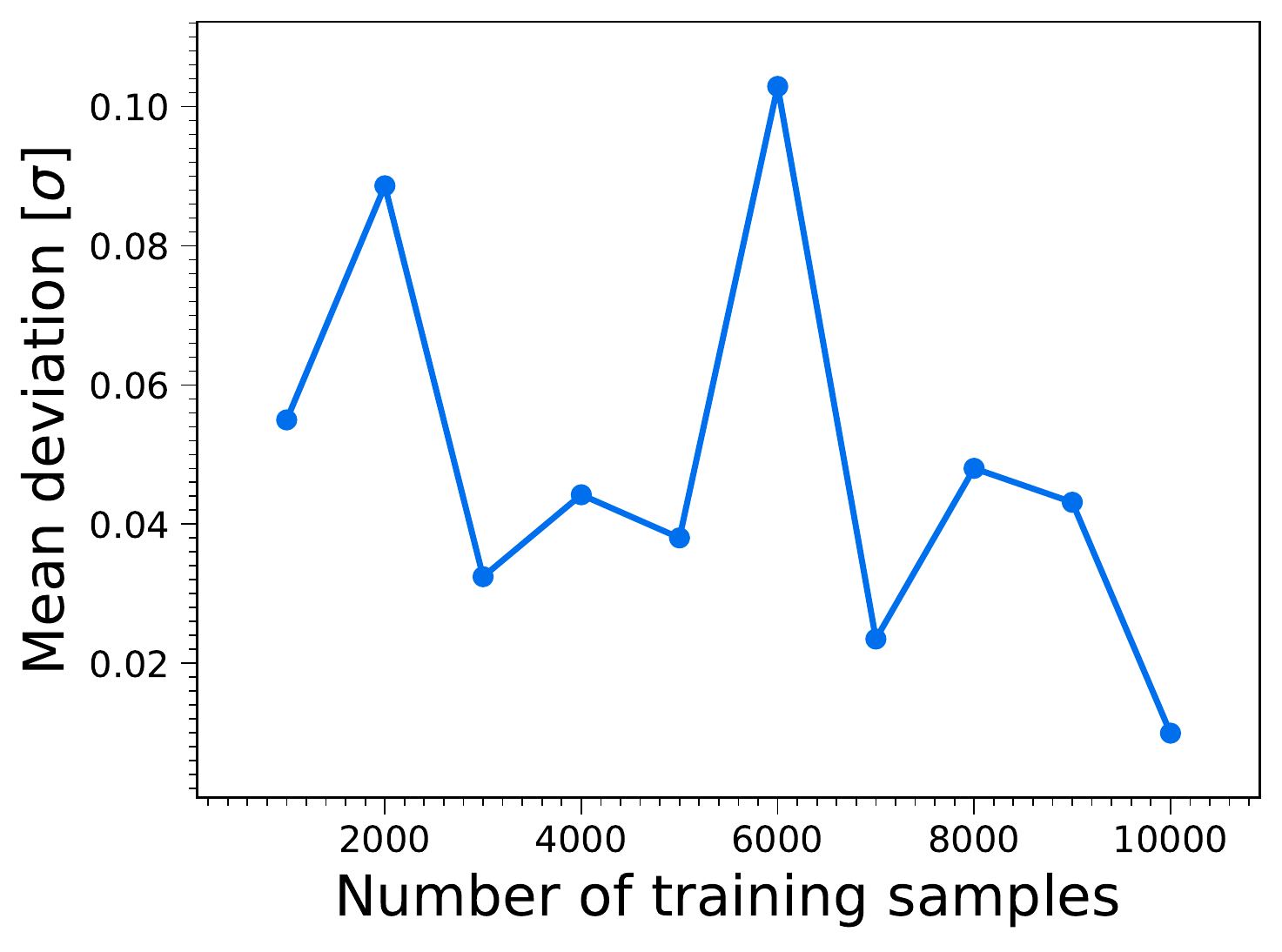}
	\includegraphics[width=0.45\textwidth]{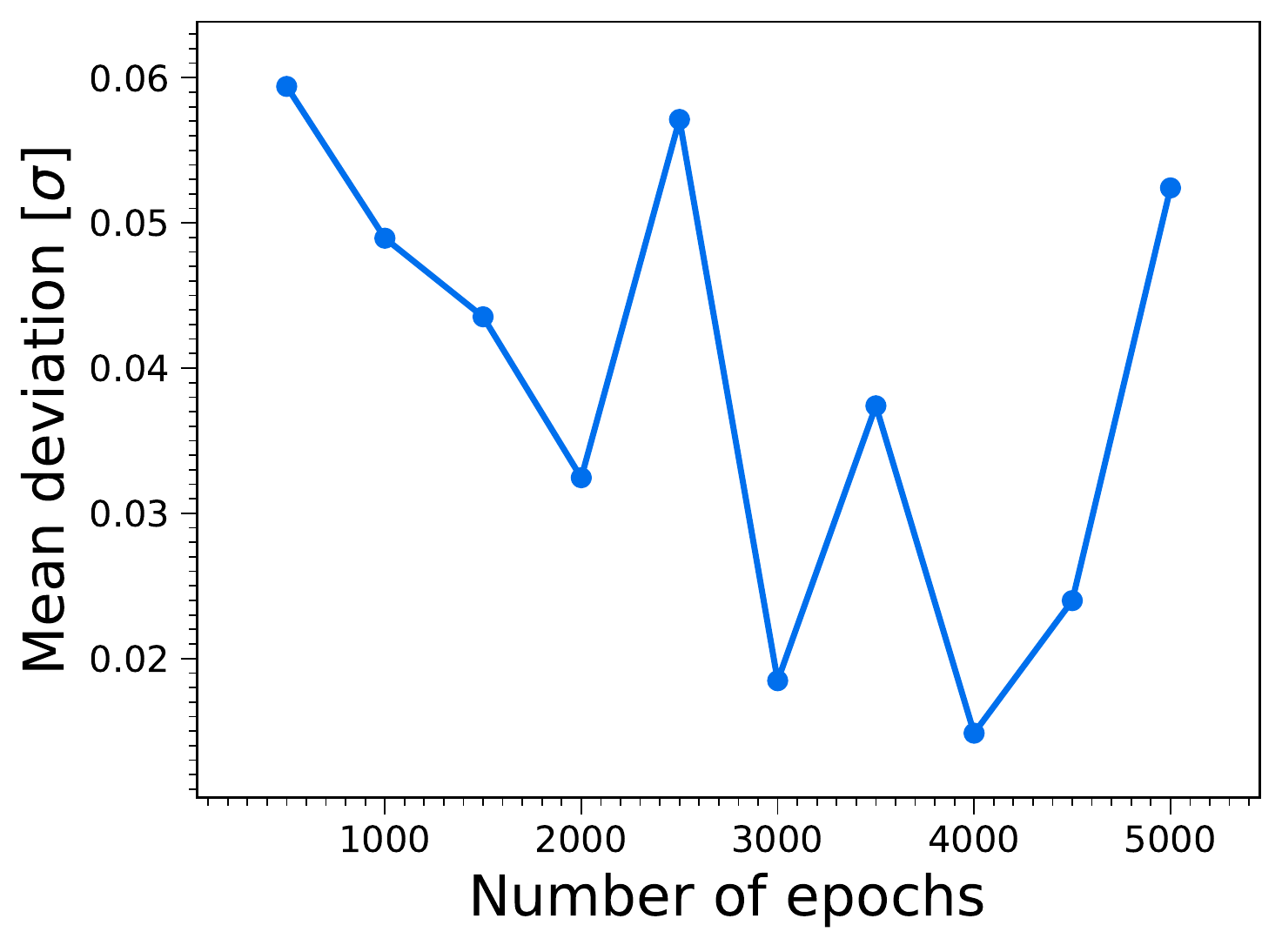}
	\caption{Mean deviations between the MDN results and the fiducial values as a function of the number of hidden layers, the number of training samples, the number of epochs, and the activation function.}\label{fig:effect_of_hyperparameters}
\end{figure*}

\subsection{The Number of Hidden Layers}\label{sec:effect_of_hiddenLayer}

To test the effect of hyperparameters in the following sections, we simulate SN-Ia data based on a future Wide-field Infrared Survey Telescope (WFIRST) experiment \citep{Spergel:2015} in a flat $w$CDM model with the following fiducial values of parameters: $w=-1$, $\Omega_{\rm m}=0.3$, and $H_0=70 ~\rm km\ s^{-1}\ Mpc^{-1}$. The total number of SN-Ia samples is 2725 in the range of $0.1<z<1.7$. We then estimate $w$ and $\Omega_{\rm m}$ using the MDN method with three Gaussian components. Finally, we calculate the mean deviation between the MDN results and the fiducial values:
\begin{equation}
{\rm Mean\ deviation}=\frac{1}{N}\left( \sum_{i=1}^N \frac{|\theta_{i, \rm pred}-\theta_{i, \rm fid}|}{\sigma_{i, \rm pred}} \right) ,
\end{equation}
where $N$ is the number of cosmological parameters, $\theta_{i, \rm fid}$ is the fiducial parameter, and $\theta_{i, \rm pred}$ and $\sigma_{i, \rm pred}$ are the predicted best-fitting value and error of the cosmological parameter, respectively.

First, we test the effect of the number of hidden layers of the MDN model. We design five different MDN structures with the number of hidden layers varying from one to five. We consider RReLU here as the activation function. We train the MDN models with 3000 samples, and after 2000 epochs, we obtain the corresponding five ANN chains by feeding the simulated observations. We then calculate the mean deviations between the MDN results and the fiducial values. The mean deviation as a function of the number of hidden layers is shown in the upper left panel of Figure~\ref{fig:effect_of_hyperparameters}. The mean deviations of these five cases are $0.072\sigma$, $0.033\sigma$, $0.032\sigma$, $0.094\sigma$, and $0.082\sigma$, respectively. We can see that there is a large deviation for MDNs with fewer or more hidden layers. An MDN with one hidden layer may not have enough layers to build a mapping between the input data and the output parameters, while an MDN with four or five hidden layers will have more difficulty learning the mixture model. We note that the training time will increase as the number of hidden layers increases. As a result, MDNs with many hidden layers become unpragmatic because of the training time.

\subsection{Activation Function}\label{sec:effect_of_activationFunction}

The settings of the activation function may also affect the parameter estimation. Here we select several commonly used activation functions and test their effects on the performance of the MDN model: rectified linear unit (ReLU), exponential linear unit (ELU), RReLU, sigmoid function, and hyperbolic tangent (Tanh).
The ReLU activation function is defined as~(see also~\citealt{Nair:2010})
\begin{equation}\label{equ:relu}
f(x)=\left\{\begin{matrix}
x & \text{if } x \geq 0 \\
0 & \text{if } x < 0 .
\end{matrix}\right.
\end{equation}
The ELU is~\citep{Clevert:2016}
\begin{equation}\label{equ:elu}
f(x)=\left\{\begin{matrix}
x & \text{if } x > 0 \\
\alpha(e^x - 1) & \text{if } x \leq 0 ,
\end{matrix}\right.
\end{equation}
where $\alpha$ is a parameter and we set it to 1. The sigmoid function is defined by (see also~\citealt{Han:1995})
\begin{equation}
f(x) = \frac{1}{1 + e^{-x}},
\end{equation}
and the hyperbolic tangent is~\citep{Malfliet:1992}
\begin{equation}
f(x) = \frac{e^x - e^{-x}}{e^x + e^{-x}}.
\end{equation}

The MDN model contains three hidden layers in our work. Using the data simulated in Section \ref{sec:effect_of_hiddenLayer}, we train another five MDN models with 3000 samples to estimate $w$ and $\Omega_{\rm m}$. After 2000 epochs, we obtain five ANN chains, and then calculate the corresponding deviations. The mean deviation of the parameters between the MDN results and the fiducial values as a function of the activation function is shown in the upper right panel of Figure~\ref{fig:effect_of_hyperparameters}. One can see that the MDN with the sigmoid function has the maximum deviation of $0.158\sigma$, while the deviations of the other MDNs are all smaller than $0.067\sigma$. Therefore, in our specific application of the MDN, the ReLU, ELU, RReLU, and Tanh activation functions can all be used for the MDN method.

\subsection{The Number of Training Samples}\label{sec:effect_of_trainingNumber}

To test the effect of the number of samples in the training set, we train several MDN models with the number of samples varying from 1000 to 10,000. The MDN model used here contains three hidden layers, the RReLU nonlinear function is taken as the activation function, and the MDN is trained with 2000 epochs. The mean deviation of parameters between the MDN results and the fiducial values is shown in the lower left panel of Figure~\ref{fig:effect_of_hyperparameters}. We can see that the mean deviation varies in the interval $[0.010\sigma, 0.103\sigma]$ as the number of training samples changes. But most of the deviations are less than $0.055\sigma$. Therefore, in our specific application of the MDN, thousands of samples can make an MDN model well trained. Note that the time of training of an MDN model is related to the number of samples in the training set. Therefore, considering the performance and training time of the MDN, the number of samples should be selected reasonably.

\subsection{The Number of Epochs}\label{sec:effect_of_epoch}

The number of epochs may also affect the performance of the MDN. To test this, we train several MDN models with the epoch varying from 500 to 5000. The MDN models used here contain three hidden layers, the activation function is RReLU, and 3000 samples are used to train the models. With the same procedure, we train and calculate deviations between the MDN results and the fiducial values. The mean deviation as a function of the number of epochs is shown in the lower right panel of Figure~\ref{fig:effect_of_hyperparameters}. We can see that as the number of epochs increases, the mean deviation oscillates in the range of $[0.015\sigma, 0.059\sigma]$. Also, in our case, there is little improvement by training for more than 2000 epochs. Since the training time is also related to the number of epochs, the number of epochs should also be selected reasonably.

\section{\bf Beta Mixture Model}\label{sec:beta_mixture_model}

\subsection{\bf One Parameter}\label{sec:beta_MDN_one_parameter}

The analysis above is focused on the Gaussian distribution as the unit of the mixture model. Since the unit of the mixture model can be any type of distribution, we use a beta distribution as the unit of the mixture model in this section. A beta mixture model with $K$ components has the form
\begin{align}\label{equ:pdf_of_beta}
\nonumber p(\theta|\bm{d}) &= \sum_{i=1}^K \omega_i {\rm Beta}(\theta; \alpha_i, \beta_i) \\
&= \sum_{i=1}^K \omega_i \cdot \frac{\Gamma(\alpha_i+\beta_i)}{\Gamma(\alpha_i)\Gamma(\beta_i)} \theta^{\alpha_i-1} (1-\theta)^{\beta_i-1},
\end{align}
where $\omega$ is the mixture weight, $\alpha, \beta>0$ are two shape parameters to be determined, and $\Gamma(\cdot)$ is the gamma function defined as
\begin{equation}
\Gamma (z) = \int_0^\infty x^{z-1}e^{-x} {\rm d}x.
\end{equation}
Here we use the Softplus function (Equation~(\ref{equ:softplus})) to guarantee $\alpha$ and $\beta$ are positive. Note that the beta distribution is defined on the interval [0, 1]; thus we normalize the parameters in the training set (i.e. the target) to the range [0, 1] by using the min$-$max normalization technique:
\begin{equation}
z = \frac{x - \text{min}(x)}{\text{max}(x) - \text{min}(x)}.
\end{equation}
The parameters of the beta mixture model can be estimated by minimizing the loss function
\begin{eqnarray}
\label{equ:loss_mdn_beta}
\mathcal{L} &=& \mathbb{E}\left[ -\ln\left( \sum_{i=1}^K \omega_i \right. \right. \nonumber \\
& \times & \left.\left.\frac{\Gamma(\alpha_i+\beta_i)}{\Gamma(\alpha_i)\Gamma(\beta_i)} \theta^{\alpha_i-1} (1-\theta)^{\beta_i-1}\right) \right] .
\end{eqnarray}
For numerical stability, we also use the Log-Sum-Exp trick; then Equation~ (\ref{equ:loss_mdn_beta}) can be rewritten as
\begin{align}
\mathcal{L} &= \mathbb{E}\left[ -\ln\left( \sum_{i=1}^K e^{\left[\ln(\omega_i) + \ln(p_i(\theta|\bm{d}))\right]}\right) \right] ~,
\end{align}
where
\begin{eqnarray}
\ln\left[p_i(\theta|\bm{d})\right] &=& \ln(\Gamma(\alpha_i+\beta_i)) - \ln(\Gamma(\alpha_i)) - \ln(\Gamma(\beta_i)) \nonumber \\
&+& (\alpha_i-1)\ln(\theta) + (\beta_i-1)\ln(1-\theta).
\end{eqnarray}

\begin{table}
	\centering
	\caption{Best-fit values of $\Omega_{\rm m}$ of the $w$CDM model using the Pantheon SN-Ia. Here $w$ is manally set to $-1$ or $-0.5$. The error is $1\sigma$ C.L. The MDN-Gaussian refers to the MDN with Gaussian components, and MDN-Beta refers to the MDN with beta components.}\label{tab:params_pantheon_cov_omm}
	\begin{tabular}{l|c|c}
		\hline\hline
		Methods & $w=-1$ & $w=-0.5$ \\
		\hline
		MCMC & $\Omega_{\rm m}=0.397\pm0.010$ & $\Omega_{\rm m}=0.0003_{-0.0002}^{+0.0151}$ \\
		\hline
		MDN-Gaussian & $\Omega_{\rm m}=0.396\pm0.009$ & $\Omega_{\rm m}=0.0033_{-0.0018}^{+0.0134}$ \\
		\hline
		MDN-Beta & $\Omega_{\rm m}=0.397\pm0.010$ & $\Omega_{\rm m}=0.0003_{-0.0002}^{+0.0150}$ \\
		\hline\hline
	\end{tabular}
\end{table}

\begin{figure}
	\centering
	\includegraphics[width=0.45\textwidth]{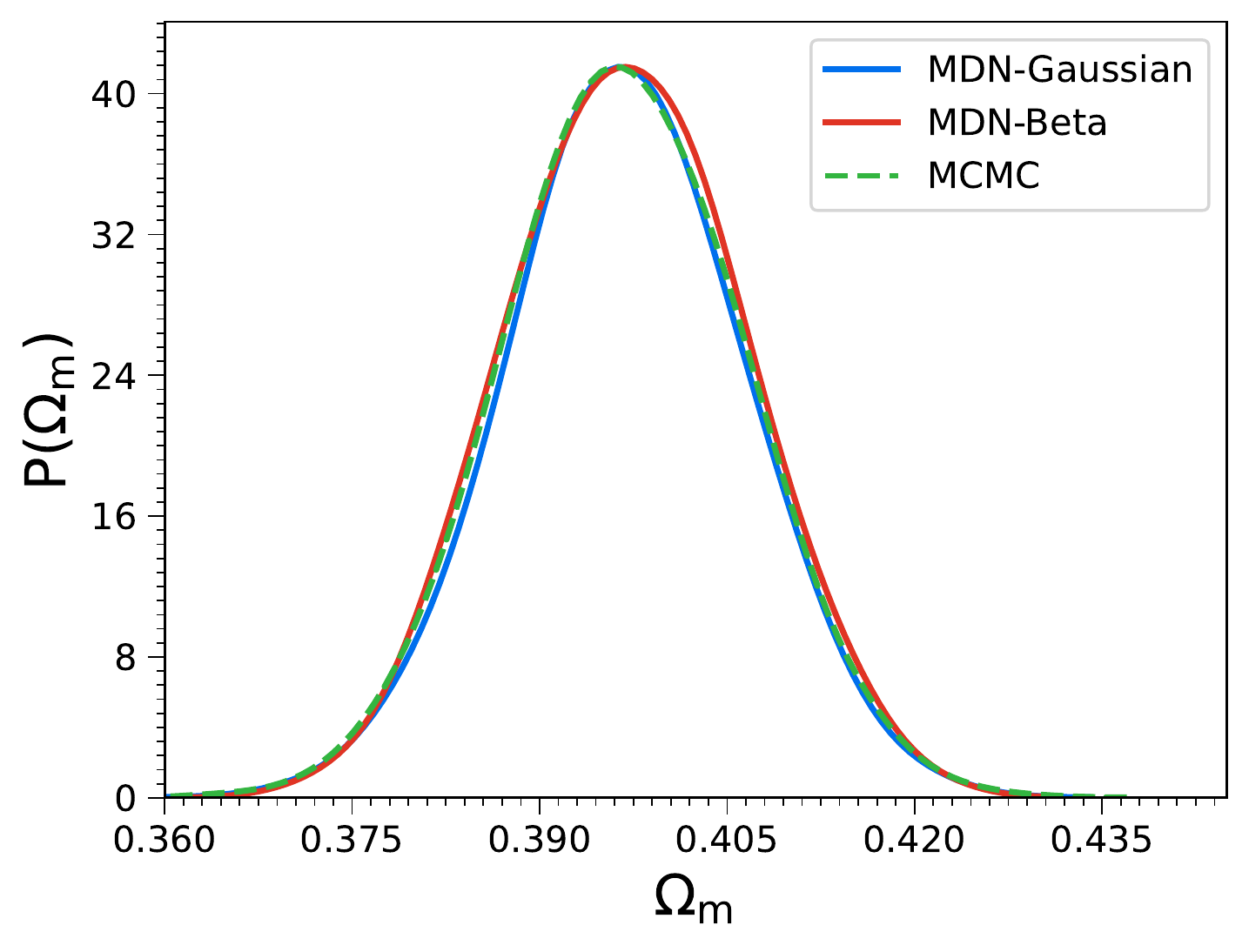}
	\caption{One-dimensional distributions of $\Omega_{\rm m}$ constrained from Pantheon SN-Ia, where the blue solid line is based on the MDN method with three Gaussian components and the red solid line is based on the MDN method with one beta component. Here $w$ is set to -1. }\label{fig:contour1D_pantheon_Gaussian_Beta_w-1}
\end{figure}
\begin{figure}
	\centering
	\includegraphics[width=0.45\textwidth]{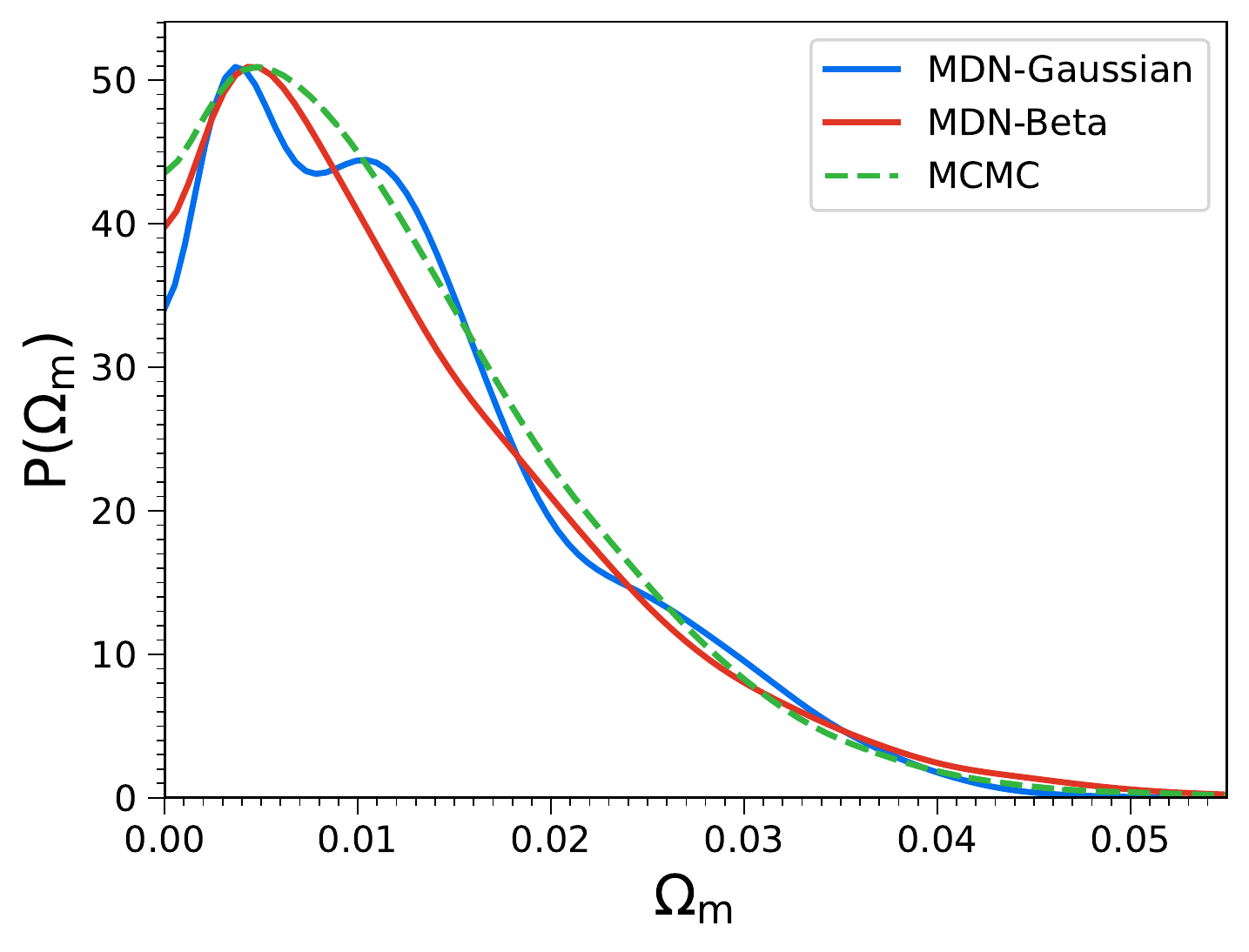}
	\caption{The same as Figure~\ref{fig:contour1D_pantheon_Gaussian_Beta_w-1}, but now $w$ is manually set to -0.5.}\label{fig:contour1D_pantheon_Gaussian_Beta_w-0.5}
\end{figure}

Since the beta distribution is for one random variable, thus to test the capability of the beta mixture model for parameter inference, we use the Pantheon SN-Ia to constrain $\Omega_{\rm m}$ of the $w$CDM model. In our analysis, the Hubble constant is set to $H_0=70 ~\rm km\ s^{-1}\ Mpc^{-1}$, and the absolute magnitude is set to $M_B=-19.3$. We consider two cases of the equation of state of dark energy: $w=-1$ and $w=-0.5$, respectively. The case $w=-0.5$ considered here is completely set manually, which is to test the case where the posterior distribution of the parameter is a truncated distribution under the physical limit. We constrain $\Omega_{\rm m}$ using the MCMC method and the MDN method. For the MDN method, we consider two mixture models: the Gaussian mixture model and the beta mixture model.

For the case of $w=-1$, we first constrain $\Omega_{\rm m}$ using the MCMC method. After generating an MCMC chain, the best-fitting value with $1\sigma$ error of $\Omega_{\rm m}$ can be calculated (see Table~\ref{tab:params_pantheon_cov_omm}). The corresponding marginalized distribution is shown in Figure~\ref{fig:contour1D_pantheon_Gaussian_Beta_w-1} with a green dashed line. We then use the MDN method with three Gaussian components to constraint $\Omega_{\rm m}$ and find the result (the blue solid line in Figure~\ref{fig:contour1D_pantheon_Gaussian_Beta_w-1}) is almost the same as that of the MCMC method for both the best-fitting value and the $1\sigma$ error. We also constrain $\Omega_{\rm m}$ using the MDN method with one beta component and obtain the same result of the MCMC method.

With the same procedure, we constrain $\Omega_{\rm m}$ for the case of $w=-0.5$. The result of the MCMC method is shown in Table~\ref{tab:params_pantheon_cov_omm} and the marginalized distribution is shown in Figure~\ref{fig:contour1D_pantheon_Gaussian_Beta_w-0.5} with a green dashed line. We can see the distribution is a truncated distribution due to the physical limit. Then, we constrain $\Omega_{\rm m}$ with the MDN method with Gaussian components and find both the best-fitting value and the $1\sigma$ error deviate slightly from the MCMC result, and the shape of the distribution (the blue solid line in Figure~\ref{fig:contour1D_pantheon_Gaussian_Beta_w-0.5}) differs slightly from that of the MCMC method. Finally, the result of $\Omega_{\rm m}$ based on the MDN method with beta components is almost the same as that of the MCMC method, and the shape of the distribution (the red solid line in Figure~\ref{fig:contour1D_pantheon_Gaussian_Beta_w-0.5}) is consistent with the one based on the MCMC method.

\begin{figure}
	\centering
	\includegraphics[width=0.45\textwidth]{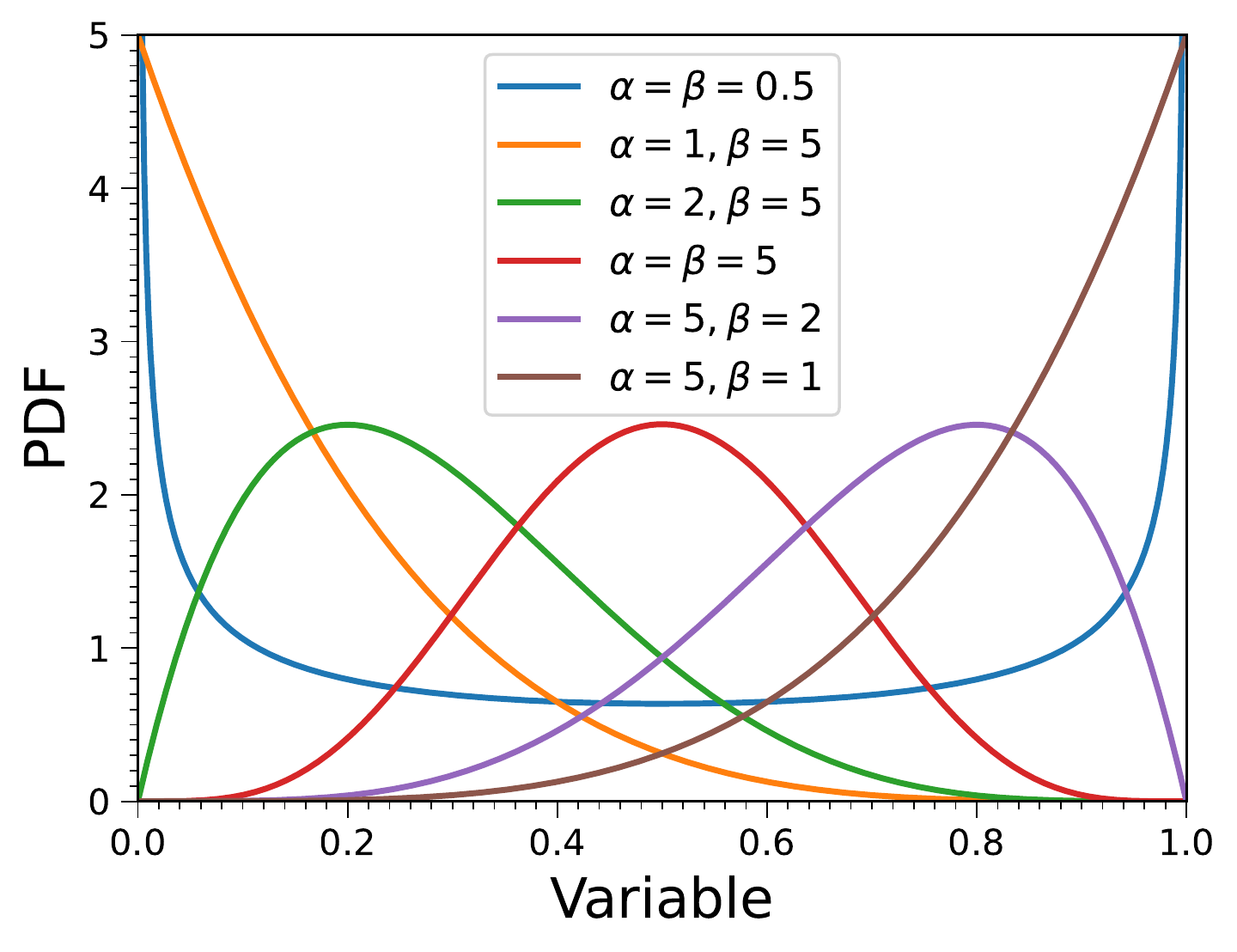}
	\caption{Examples of beta probability density function with different shape parameters calculated using the beta function in Equation~(\ref{equ:pdf_of_beta}).}\label{fig:beta_distribution_example}
\end{figure}

These results show that MDN models with beta components are able to constrain parameters, and even have advantages over MDN models with Gaussian components in the case of truncated distribution. In the case of truncated distribution, the reason for the beta mixture model performing better than the Gaussian mixture model is that the beta distribution has more shape features (see Figure~\ref{fig:beta_distribution_example} for examples), which enables it to fit better for the posterior distribution of parameters.

\begin{figure*}
	\centering
	\includegraphics[width=0.96\textwidth]{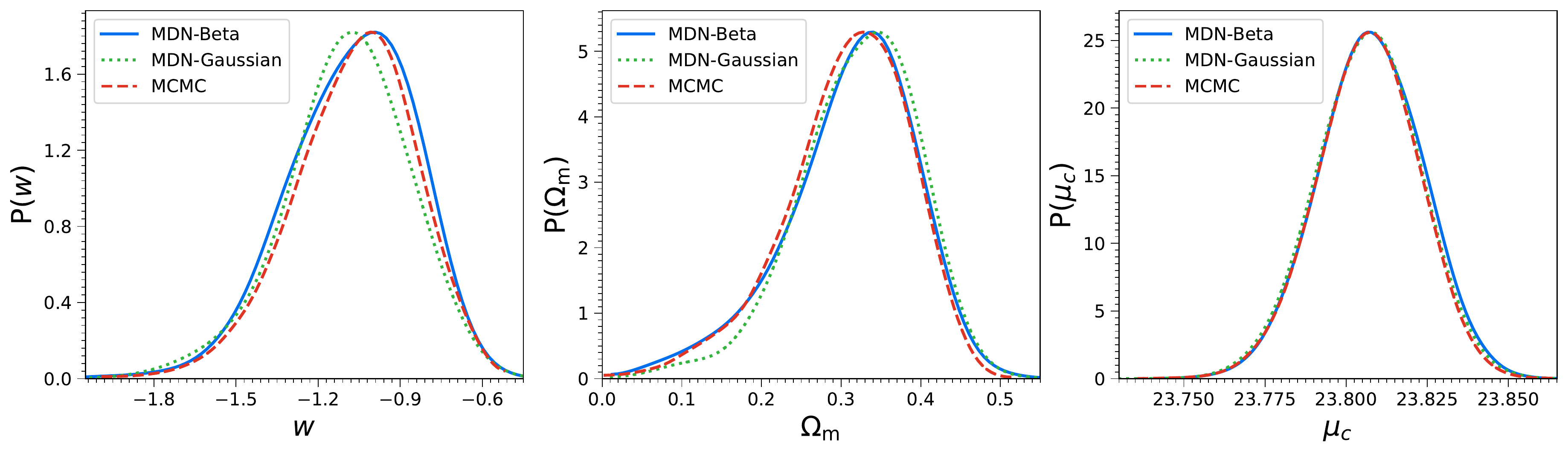}
	\caption{One-dimensional distributions of $w$, $\Omega_{\rm m}$, and $\mu_c$ constrained from Pantheon SN-Ia.}\label{fig:contour1D_pantheon_Beta}
\end{figure*}

\subsection{\bf Multiple Parameters}\label{sec:beta_MDN_multiple_parameters}

We now switch to multiparameter estimation. If parameters are not independent, the beta mixture model cannot be used when doing the MDN parameter estimation. But if the correlations between parameters are negligible, the beta mixture model can be used for multiple-parameter tasks to obtain the one-dimensional distribution of each parameter. In this case, the probability density in Equation~(\ref{equ:pdf_of_beta}) becomes
\begin{align}\label{equ:pdf_of_beta_independent}
p(\bm\theta|\bm{d}) &= \prod_{j=1}^N \sum_{i=1}^K \omega_i \cdot \frac{\Gamma(\alpha_i+\beta_i)}{\Gamma(\alpha_i)\Gamma(\beta_i)} \theta_j^{\alpha_i-1} (1-\theta_j)^{\beta_i-1},
\end{align}
where $N$ is the number of parameters. To test this, we constrain $w$, $\Omega_{\rm m}$, and the nuisance parameter $\mu_c$ using the Pantheon SN-Ia. The Softplus function is used as the activation function considering the stability of the MDN. The one-dimensional distributions of these parameters are shown in Figure~\ref{fig:contour1D_pantheon_Beta}. We can see that the results based on the MDN with beta components are almost the same as those of the MCMC method. Our analysis shows the instability of the MDN when using Equation (\ref{equ:pdf_of_beta_independent}) as the probability density. The reason may be that there are correlations between parameters, but they are not considered by the MDN. Besides, there are too many components ($N\times K$ components in total) for the probability density in Equation (\ref{equ:pdf_of_beta_independent}), especially for cases of many cosmological parameters, which will also increase the instability of the MDN. Therefore, the beta mixture model based on Equation (\ref{equ:pdf_of_beta_independent}) should be further investigated.
 
As a comparison, we also constrain $w$, $\Omega_{\rm m}$, and $\mu_c$ without considering correlations between them, by using an MDN with a Gaussian mixture model. In this case, the probability density in Equation (\ref{equ:pdf_of_gaussian_1}) becomes
\begin{align}
p(\bm\theta|\bm{d}) &= \prod_{j=1}^N \sum_{i=1}^K \omega_i\cdot\frac{1}{\sqrt{2\pi\sigma^2_i}}e^{-\frac{(\theta_j-\mu_i)^2}{2\sigma^2_i}}~.
\end{align}
The one-dimensional distributions of these three parameters are shown as green dotted lines in Figure~\ref{fig:contour1D_pantheon_Beta}. Obviously, this result is similar to that based on the beta mixture model. We note that the Gaussian mixture model performs poorly for parameters with truncated distribution (see Figure \ref{fig:contour1D_pantheon_Gaussian_Beta_w-0.5}), and correlations between parameters will not be obtained. Therefore, the Gaussian mixture model based on Equation~(\ref{equ:pdf_of_gaussian_multi}) is recommended for multiparameter estimates.

\section{\bf Discussion}\label{sec:discussions}

The MDN method proposed in this paper combines an ANN and a mixture model. The main idea is to assume that the posterior distribution of parameters is a mixture of several elemental simple distributions, such as the Gaussian distribution. Once the basic distribution is determined, the posterior distribution of the parameters can be calculated. Therefore, the neural network here aims to determine the basic distribution. In general, the MDN method models a connection between measurements and parameters, where the function of the mixture model is to model the parameters, and the function of the neural network is to extract information from the measurement. The neural network used here is a fully connected network that can deal with one-dimensional data. In future work, we will generalize this technique to two- and three-dimensional cosmological data sets.

As we illustrated in Section \ref{sec:training_parameter_estimation} that the parameter space is updated in the training process, the parameter space will gradually approach and cover the actual values of the cosmological parameters. Therefore, the MDN method is not sensitive to the initial conditions, suggesting that the initial conditions can be set freely with an arbitrary range of parameters. This fact is an advantage over the MCMC method and is beneficial for models that lack prior knowledge.

Many hyperparameters can be chosen when using the MDN, such as the number of hidden layers, the activation function, the number of training samples, the number of epochs, and even the number of components. The analysis in Section \ref{sec:effect_of_hyperparameters} uses specific simulated (or observational) data to test the effect of these hyperparameters. Due to the selection of hyperparameters and randomization in the training process (e.g. the initialization of the network parameters ($w$, $b$) in Equation~(\ref{equ:neuron_function}) and the random noise added to the training set), the final estimated parameters have certain instability. But this shortcoming does not affect the fact that the MDN can get an accurate estimation with a mean deviation level of $\mathcal{O}(10^{-2}\sigma)$, after selecting appropriate hyperparameters (e.g. using thousands of samples and thousands of epochs).

\section{\bf Conclusions}\label{sec:conclusions}
In this work, we conducted parameter inference with the MDN method on both one data set and multiple data sets to achieve joint constraints on parameters using the Pantheon SN-Ia and the CMB power spectra of the {\it Planck} mission. Our analysis shows that a well-trained MDN model can estimate parameters at the same level of accuracy as the traditional MCMC method. Furthermore, compared to the MCMC method, the MDN method can provide accurate parameter estimates with far fewer forward simulation samples.

We considered two mixture models in the MDN model: the Gaussian mixture model and the beta mixture model. The Gaussian mixture model can theoretically be applied to any parameter, while the beta mixture model is only applicable to a case with one parameter or multiple independent parameters. Our results show that the beta mixture model outperforms the Gaussian mixture model for parameters with truncated distributions.

The MDN is an ANN-based method, making it a flexible method that is extendable to two- and three-dimensional data, which may be of great use in extracting information from high-dimensional data.

\section{\bf Acknowledgement}
Y.-Z.M. acknowledges the support of the National Research Foundation with grant Nos. 120385 and 120378. J.-Q.X. is supported by the National Science Foundation of China under grant No. U1931202. All computations were carried out using the computational cluster resources at the Centre for High Performance Computing, Cape Town, South Africa. This work was part of the research program ``New Insights into Astrophysics and Cosmology with Theoretical Models Confronting Observational Data'' of the National Institute for Theoretical and Computational Sciences of South Africa.

\end{document}